\documentclass[review,1p,authoryear]{elsarticle}

\usepackage{amssymb}
\usepackage{graphics}
\usepackage{gensymb}
\usepackage{color}
\usepackage{etoolbox}
\usepackage[draft]{todonotes}
\usepackage{subfig}
\usepackage{lipsum}
\usepackage{tikz}
\usepackage{pgfplots}
\usepackage{verbatim}
\usepackage{float}
\usepackage{mwe,tikz}
\usepackage[percent]{overpic}
\usepackage{verbatim}
\usepackage{booktabs}
\setcitestyle{numbers}
\usepackage{setspace}
\usepackage{times}
\newcommand{\EN}{\mathrm{(\textit{E/N})_{eff}}}
\setcitestyle{square,comma}

\journal{Journal of Physics D}

\begin{document}

\begin{frontmatter}



\title{Characteristics of a novel nanosecond DBD microplasma reactor for flow applications}


\author[label1]{A. Elkholy} \ead{A.elkholy@tue.nl}
\author[label2]{E. van Veldhuizen}
\author[label2]{S. Nijdam}
\author[label2]{U. Ebert}
\author[label1]{J. van Oijen}
\author[label1]{N. Dam}
\author[label1]{L. Philip H. de Goey}

\address[label1]{Department of Mechanical Engineering, Technische Universiteit Eindhoven, Netherlands.}
\address[label2]{Department of Applied Physics, Technische Universiteit Eindhoven, Netherlands.}

\begin{abstract}
 We present a novel microplasma flow reactor using a dielectric barrier discharge (DBD) driven by repetitively nanosecond high-voltage pulses. Our DBD-based geometry can generate a non-thermal plasma discharge at atmospheric pressure and below in a regular pattern of micro-channels. This reactor can work continuously up to about 100 minutes in air, depending on pulse repetition rate and operating pressure. We here present the geometry and the main characteristics of the reactor. Pulse energies of 1.9 $\mu$J and 2.7 $\mu$J per channel at atmospheric pressure and 50 mbar, respectively, have been determined by time-resolved measurements of current and voltage. Time-resolved optical emission spectroscopy measurements have been performed to calculate the relative species concentrations and temperatures (vibrational and rotational) of the discharge. Effects of the operating pressure and the flow velocity on the discharge intensity have been investigated. In addition, the effective reduced electric field strength $\EN$ has been obtained from the intensity ratio of vibronic emission bands of molecular nitrogen at different operating pressures. The derived $\EN$ increases gradually from 500 Td to 600 Td when decreasing the pressure from one bar to 0.4 bar. Below 0.4 bar, further pressure reduction results in a significant increase in the $\EN$ up to about 2000 Td at 50 mbar.
\end{abstract}

\end{frontmatter}


\section{Introduction}
\label{intro}

Plasma flow reactors are can be used to chemically process a gas. They are of interest because of their high energy efficiency and high selectivity. Therefore, they have a strong potential in applications like surface treatment  \cite{Surface_Treatment}, thin film deposition \cite{thin_film}, detoxification of gaseous pollution \citep{pollution_control,shimizu2008study}, plasma medicine \cite{plasma_medicine} and ozone generation \cite{Ozone}. Many different designs have been tried to generate a stable non-thermal plasma discharge in a flow reactor at pressures up to one atmosphere. Some of these designs are plasma jets \cite{kolb2008cold,schutze1998atmospheric}, packed-bed plasma reactors \cite{chen2008review,yu2012characteristics}, pulsed-corona plasma reactors \citep{grabowski2007breakdown,ono2003dynamics,winands2006industrial} and atmospheric pressure glow discharges (APG) \cite{akitsu2005plasma}.

One of the most extensively studied configurations is an array of so-called microhollow cathode discharges (MHCD) which has the ability to produce a high-density plasma discharge in a relatively small volume (characteristic dimension of about 10$^{-4}$ m) \citep{gomes2009characterization,HPHCD,schoenbach1996microhollow}. Based on that, compared to other types of flow reactors, MHCD ensures a maximum interaction between the plasma discharge and the processed gas. However, two conditions should be fulfilled to maintain a discharge in MHCD devices. Firstly, the applied voltage should exceed the breakdown voltage and be able to ignite the discharge for the given gas and pressure \cite{Foest200687}. Secondly, according to the Allis-White similarity law, the product of the pressure ($p$) and the aperture diameter of the cathode ($D$) should fall in the typical operating range for MHCD, which is 0.1-13 mbar$\cdot$cm depending on gas, electrodes, and geometry \cite{becker2006microplasmas}. These conditions pose some constraints regarding scaling-up the diameter of the MHCD devices at high pressures. Based on the upper limit of the product ($p\cdot D$), atmospheric pressure operation in air would lead to a maximum diameter ($D$) of 100 $\mu$m. However, in many applications, especially for gas treatment, it is desirable to increase the diameter to reduce the pressure drop for the same flow condition.

One of the most successful ways to increase the hole diameter of the reactor is by using a dielectric barrier discharge (DBD) \cite{p2011microplasmas, sakai2005integrated} driven by nanosecond high-voltage pulses. The surface discharge deposited by the (pulsed) discharge counteracts the applied field, thereby self-limiting the discharge. Together with the nanosecond pulses, this helps to generate a uniform discharge distribution and reduces the chance of glow to arc transition. In addition, using nanosecond repetitive high-voltage pulses provides highly energetic electrons and more chemically excited species compared to AC power sources due to the rapid ionization process, with a power consumption that is about one order of magnitude lower \cite{zhang2013comparison,pai2009nanosecond}. The increase of the electron energy comes without a considerable increase of the gas temperature, thereby reducing undesired thermal effects of the plasma discharge on the treated gas.

In this paper, we introduce a new geometry which utilizes the dielectric barrier discharge (DBD) method to sustain a non-thermal plasma discharge in an array of 400 $\mu$m diameter channels perforating a 1.5 mm thick dielectric slab with embedded electrodes. The discharge is powered by nanosecond high-voltage pulses and is operated at pressures up to one atmosphere, which corresponds to a pressure ($p$) times diameter ($D$) of 40 mbar$\cdot$cm. The main development in this study is that the high-voltage electrode is fully embedded within the dielectric substrate. This allows to generate a non-thermal plasma discharge at atmospheric pressure in channels larger than the maximum possible diameter for microhollow cathode discharges (MHCD) at the same pressure.

\begin{figure}[h]
 \centering
 \includegraphics[width=160pt]{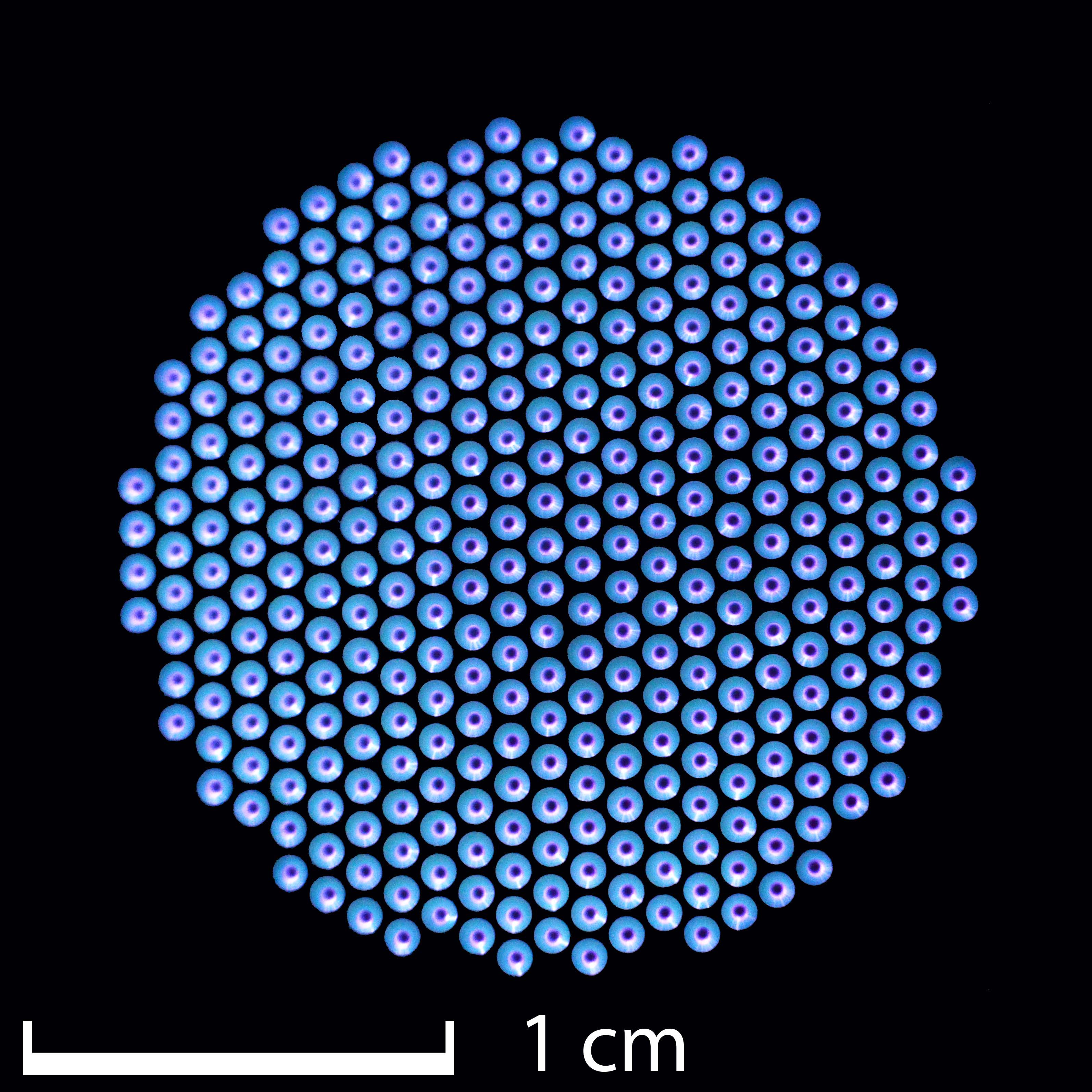}
  \caption{\small Image of the plasma discharge in the DBD microplasma reactor (see Fig. \ref{fig:Reactor} for details) in air with a flow rate of 3.4 l/min at atmospheric pressure and a pulse repetition frequency of 3 kHz, exposure time: 1/20 sec.}	
  \label{plasmadisk}
\end{figure}

The geometry consists of 363 channels of 400 $\mu$m in diameter placed in parallel. Each channel has two sets of electrodes, stacked to work (optionally) in series to increase the discharge volume in order to increase the efficiency of the gas-discharge interaction. Fig. \ref{plasmadisk} shows a natural luminosity top view image of a plasma discharge in this DBD microplasma reactor, with 3.4 l/min air flow upwards, out of the plane of the paper, at atmospheric pressure, 4 kV voltage pulse with a repetition rate of 3 kHz, recorded by a common DSLR camera (Nikon D5100).

The non-thermal nature and the high electron energy of the produced discharge make this geometry very promising for applications such as plasma-assisted combustion, chemical treatment, light sources, plasma medicine, pollutant control, and many others.

In this paper we present the geometry, material and fabrication process of the DBD microplasma reactor as well as the characteristics of the nanosecond pulsed plasma produced by it. Time-resolved optical spectroscopy is used to investigate the discharge emission and its temperatures. ICCD images are analyzed to identify the discharge structures at different operating pressures. The relation between effective reduced electric field strength $\EN$ and operating pressure and the effect of flow velocity on discharge emission are also presented. Finally, reactor lifetime is determined for different operating pressures and pulse repetition rates.

\section{Experimental Setup}

\label{sec:Experimental Setup}

\subsection{Reactor geometry, materials and fabrication}
\label{sec:device materials }

\begin{figure*}[t!]
 \hspace{-1cm}
  \includegraphics[width=450pt]{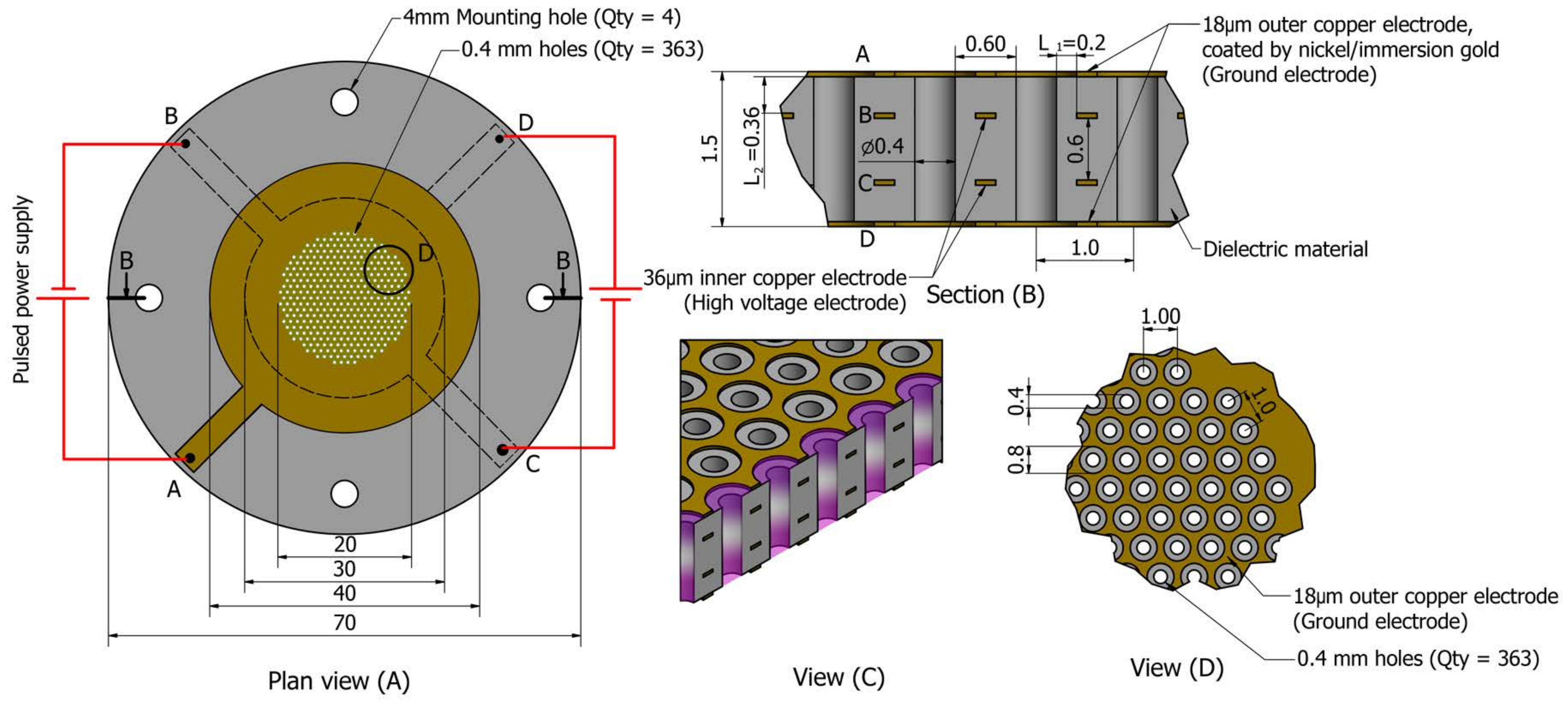}
  \caption{Schematic of reactor configuration. All dimensions are in mm. }
  \label{fig:Reactor}
\end{figure*}

As shown in Fig. \ref{fig:Reactor}, the new plasma device is a 1.5 mm thick dielectric slab perforated by a hexagonal array of 363 channels. The channels have a diameter of 0.4 mm and a pitch of 1.0 mm. The outer diameter of the reactor is 70 mm while the effective diameter of the flow area is 20 mm. The reactor geometry includes four layers of copper electrodes (A, B, C and D) in Fig. \ref{fig:Reactor}B. The high-voltage electrodes (B and C) are embedded into the dielectric material. The grounded electrodes (A and D) are located on the top and bottom surfaces of the reactor. The outer and the inner copper layers have a diameter of 40 mm and 30 mm, respectively. All electrodes are perforated in the same hexagonal pattern as the dielectric, but the holes are twice as large as those in the dielectric. The device can work in two modes: single-layer discharge mode and double-layer discharge mode. In the single-layer discharge mode only electrodes A and B are activated (see Fig. \ref{fig:circuit} for the equivalent circuit diagram) while electrodes C and D are kept floating. All the work presented in this study has been done in the single-layer discharge mode.

Cathode and anode are separated by a dielectric layer with a thickness of 0.36 mm. The outer (grounded) and inner (high-voltage) electrodes are made of copper with thicknesses of 18 $\mu$m and 38 $\mu$m, respectively. The separation distance between the two embedded electrodes is 0.6 mm, while the total thickness of the reactor is 1.5 mm. The width and thickness of the embedded electrodes have been calculated to safely withstand the peak current of the discharge pulses which is 40 A as shown in Fig. \ref{fig:pulseVI} for a duration of 10 ns.

We came to this design by performing a geometrical parameter study to i) determine the minimum dielectric thickness in order to maximize the discharge strength and ii) guide most of the discharge to the inside of the holes instead of the top or bottom reactor surfaces, to get an efficient interaction with the flowing gas. From this study we found that the dielectric thickness between the anode and the wall of the hole, ($L_1$ mm in view (B) of Fig. \ref{fig:Reactor}), should be less than the axial dielectric thickness between the anode and cathode ($L_2$).

The dielectric material of the reactor is a composite of woven electrical grade fiberglass and epoxy resin. It has a dielectric constant $D_{r}$ of 4.17 $\pm $ 0.05 (at 1 GHz / 23 $^{\circ}$C), a thermal conductivity of 0.4 W/m.K, a dielectric breakdown strength of 32 kV/mm, a thermal expansion coefficient of 14 ppm/$^{\circ}$C and a glass transition temperature of 180 $^{\circ}$C. This material has a stable performance over a wide range of frequencies and temperatures, as well as good machinability.

\begin{figure}[h]
		\hspace{3cm}
     \includegraphics[width=10cm]{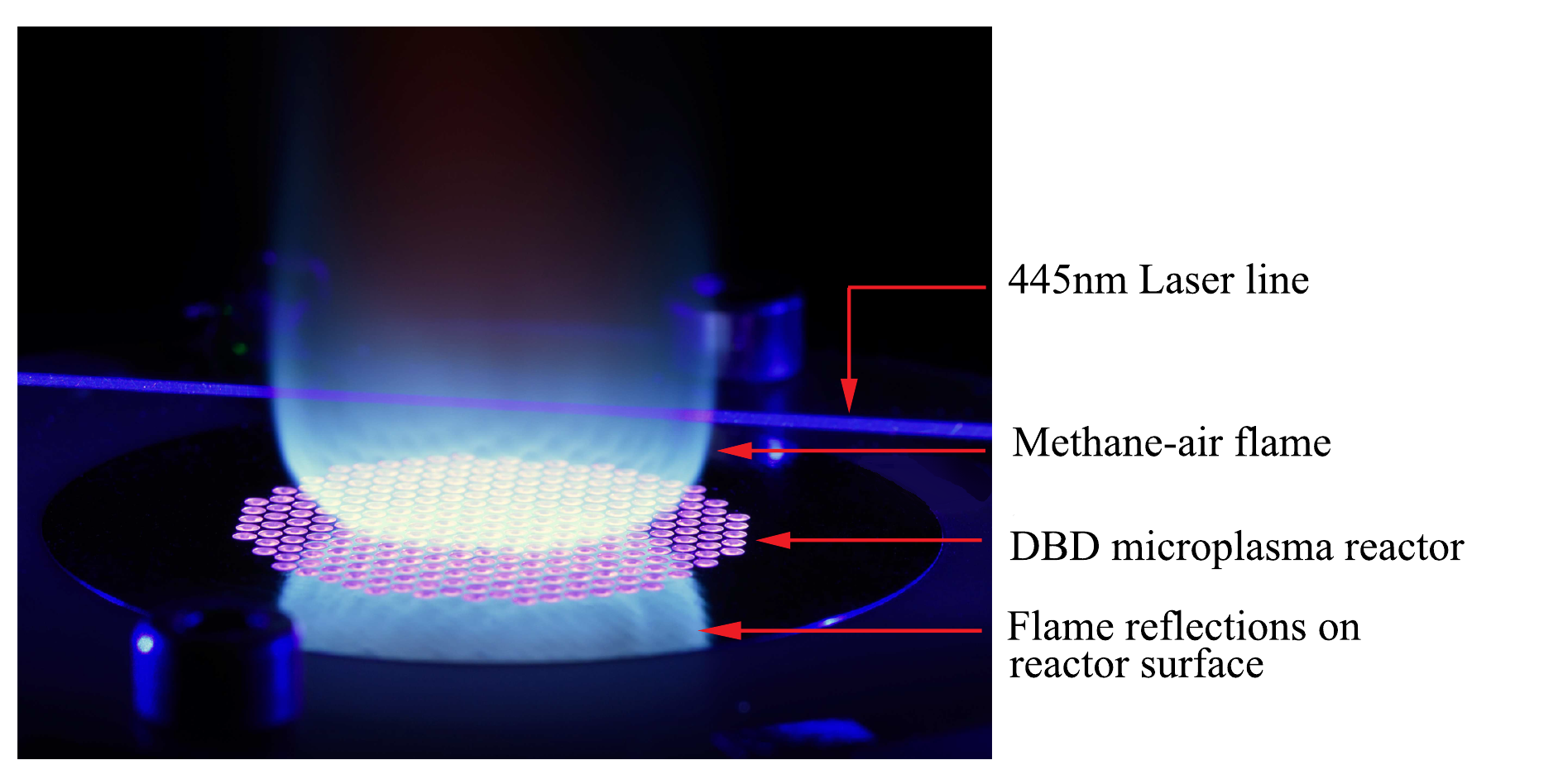}
\caption{Image of a methane-air flame stabilized on the DBD microplasma reactor at atmospheric pressure. The laser is used for flame diagnostics and is not discussed in this paper.}
 \label{fig:burner}
\end{figure}

In order to reduce the chances of micro-filamentary discharges on the reactor surface, which would reduce the reactor lifetime as a result of excessive heat, the outer copper electrodes have been coated by a nickel/immersion gold layer which gives a surface roughness of 0.5 $\mu$m. In some applications which are accompanied by thermal radiation, like plasma-assisted combustion or chemical vapour deposition, the copper layers also act as protective layers to conduct away excessive heat from the process. This prevents overheating of the dielectric material which can lead to dielectric failure. An example of the use of our reactor is given in Fig. \ref{fig:burner} where it stabilizes a methane-air flame in order to study the effect of plasma discharges on combustion characteristics at atmospheric pressure. Results of this combustion study will be published in a separate work.

The plasma reactor is manufactured by a multi-step UV lithography process. Firstly, the desired copper patterns are precisely defined on two dielectric slabs with copper layers pre-bonded onto each side. These correspond to the outer and inner electrode pairs and will be stacked on top of each other later in the process. Then, UV light is used to harden a photo-resist layer on the desired copper tracks, after which an alkaline solution is sprayed over the panel to etch away the undesired copper. Next, the two panels are stacked together bonded by another epoxy resin layer and thermally pressed to guarantee a permanent bond without any air gap. Then, the copper tracks on the upper and lower layers are coated with a nickel/immersion gold layer. Finally, the channels are drilled by a laser drilling machine to ensure clean hole walls and a uniform pattern.

\begin{figure*}[h]
\centering
	\includegraphics[width=300pt]{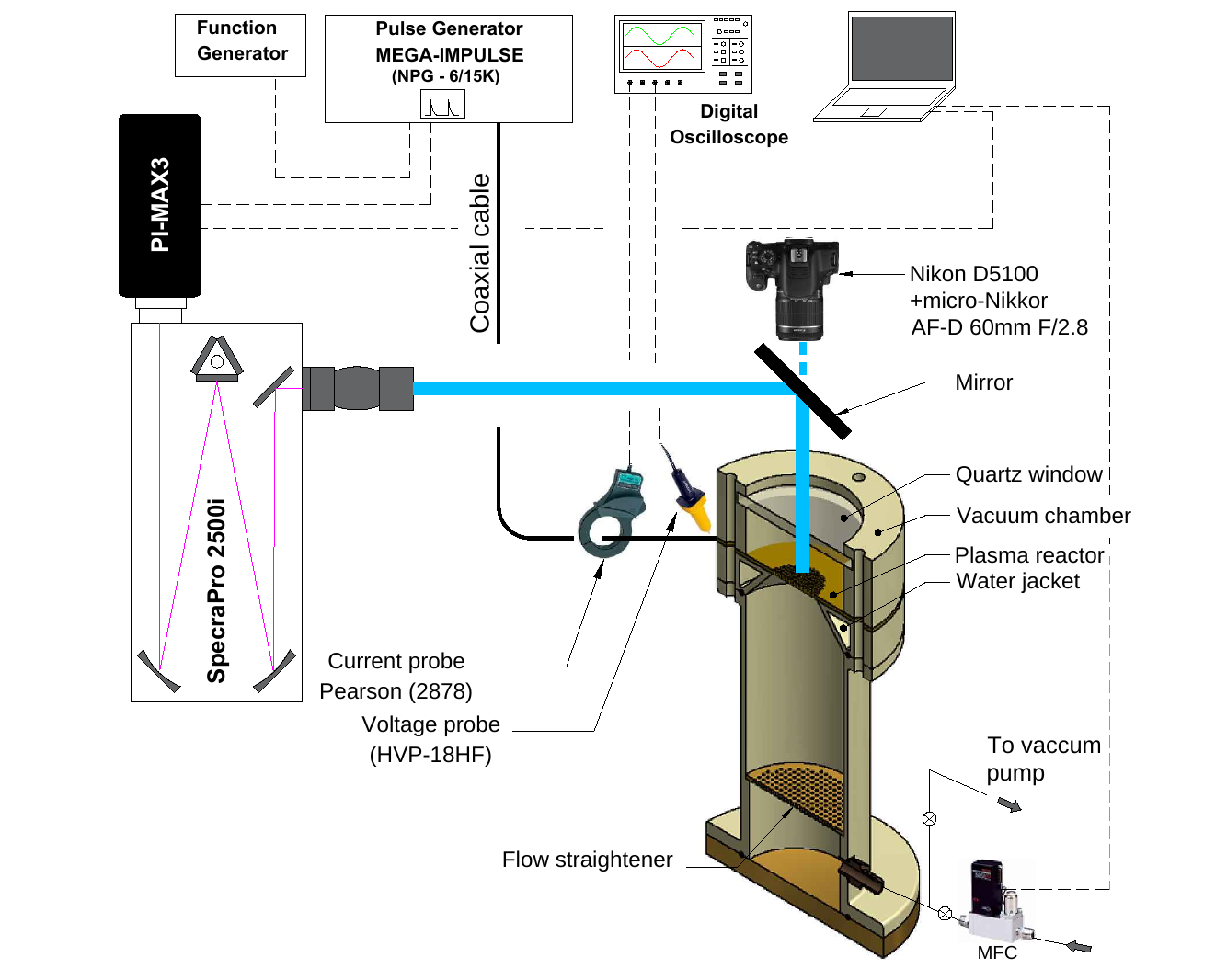}
    \caption{Schematic of the experimental setup.}
    \label{fig:testrig}
\end{figure*}


\subsection{Flow setup}
\label{Reactor geometry}

\label{Subsection:flow setup}
A schematic representation of the experimental setup is shown in Fig. \ref{fig:testrig}. The setup consists of three main parts; the plenum chamber at the bottom, a water jacket in the middle and the microplasma reactor on top. The plenum chamber and the water jacket are made of steel and brass respectively. The purpose of these parts is to create a uniform air flow towards the plasma reactor. The gas inlet port is located at the bottom of the plenum chamber. A perforated plate at 20 mm from the bottom works as a flow straightener. The holes in the perforated plate (1.0 mm in diameter and 2.0 mm in pitch) cover the entire area of the plate. In case of reduced pressure, a vacuum chamber of 20 mm height and 60 mm in diameter is mounted on top of the plasma reactor. The top flange of the chamber includes a fused quartz window with a thickness of 4 mm, through which the discharge emission can be viewed. The reactor temperature is controlled by means of the water jacket underneath the reactor. The water inlet and outlet are connected to a heater and thermostat to control the water temperature. This water circuit can heat up and control the upper surface of the reactor up to 80 \degree C.

The air flow is controlled by a mass flow controller (MFC; Bronkhorst model F-202CV). A cylindrical buffer vessel is placed before the MFC to damp pressure fluctuations. In addition, an air filter is installed just before the MFC to avoid contaminations. To ensure purity, synthetic air of grade 4.0 has been used. A vacuum pump model Busch R 5 KB 0010 has been used to evacuate the vacuum chamber to the desired pressure.

\subsection{Electrical characteristics}
\label{Pulses and energy measurements}

\begin{figure}[h]
 \centering
 \includegraphics[width=250pt]{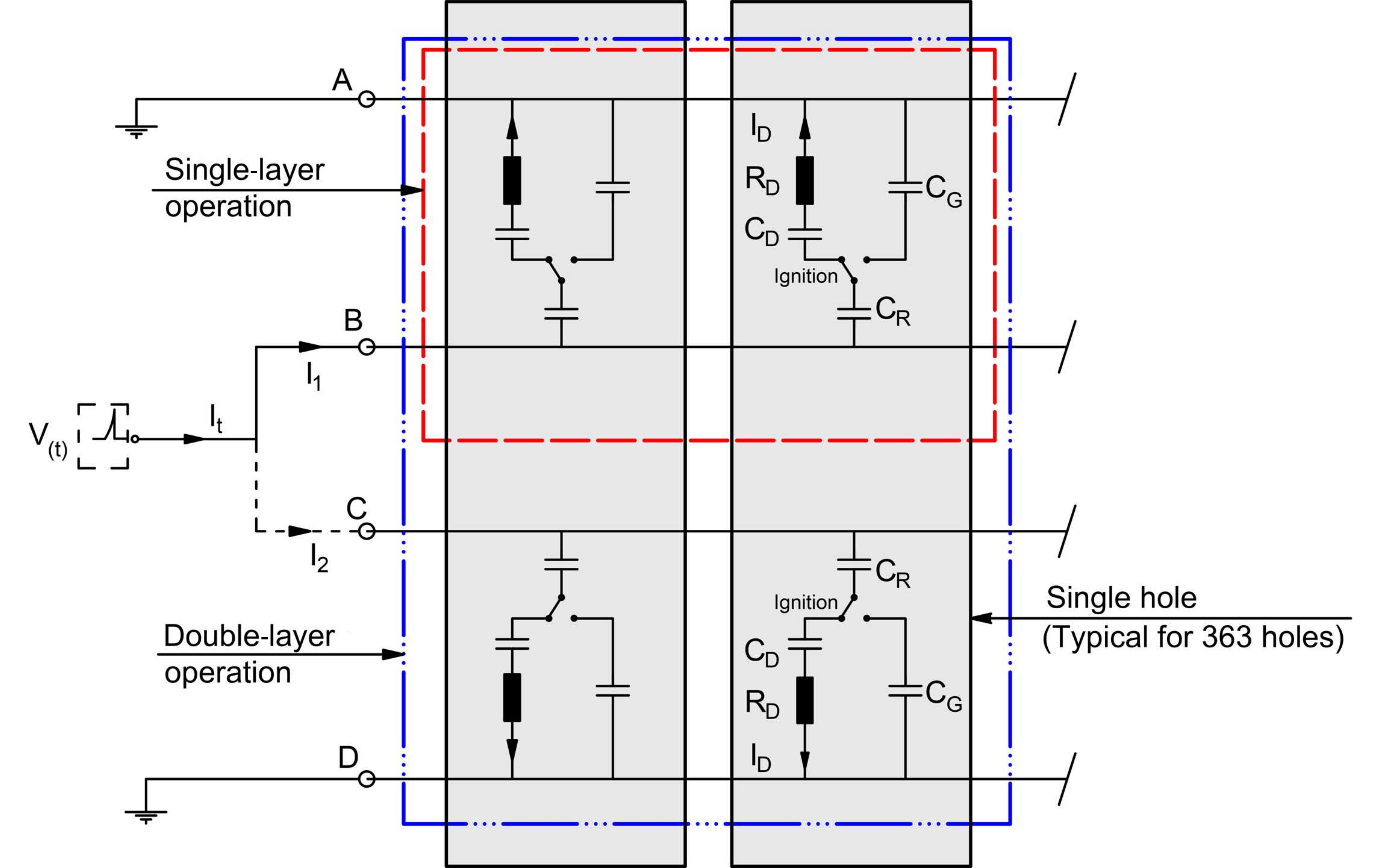}
 \caption{Equivalent electrical circuit diagram for single-layer operation (red box) and double-layer operation (blue box). For connections A, B, C and D, refer to Fig. \ref{fig:Reactor}. $C_R$ is the reactor capacitance, $C_G$ is the air gap capacitance, $R_D$ and $C_D$ are the resistance and capacitance of the discharge, respectively. For illustration purposes only 2 of the 363 holes are drawn.}
 \label{fig:circuit}
\end{figure}

Fig. \ref{fig:circuit} shows the equivalent electrical circuit diagram as used with single-layer operation (red box) and double-layer operation (blue box) for the DBD microplasma reactor. In case of double-layer operation, the total voltage $V_{(t)}$ is the same for the two layers, while the current is divided evenly between the two layers. When the voltage exceeds the breakdown voltage, indicated as an ignition switch in the figure, the current $I_D$ will pass through the discharge gap to initiate the plasma discharge.
	
The high-voltage pulses are produced by a Mega-Impulse semiconductor-based pulse generator, model NPG-6/15k. Positive polarity pulses of 4 - 6 kV in amplitude, 25 ns in duration and 10 ns rise time are created at a pulse repetition frequency in the range of 0-10 kHz. The pulse generator is externally triggered by a FeelTech frequency generator model FY2102S. Voltage and current are simultaneously monitored by a  Lecroy waverunner 44MXi-A oscilloscope with an analog bandwidth of 400 MHz. The voltage is recorded by a Pintek model HVP-18HF high-voltage probe with a sensitivity of 1 V/1 kV, a bandwidth of 150 MHz and a 2.4 ns rise time. The current through the electrodes is measured by a Pearson coil model 2878 with a 5 ns rise time and a sensitivity of 100 mV/A voltage-to-current conversion. The voltage and current signals are averaged over 100 pulses on the oscilloscope.

The total effective capacitance $C_{total}$ is the sum of two components, the reactor capacitance $C_R$ and the cable capacitance $C_{cable}$. Both the reactor and cable capacitances have been measured by a Fluke RCL meter model PM6303, resulting in $C_R=94\; pF$ and $C_{cable}=200\; pF$. So, the equivalent total capacitance is $C_{total}=64\; pF$.

\subsection{Optical emission spectroscopy setup}
\label{OES setup}
Optical emission spectroscopy has been performed with a SpectraPro 2500i spectrometer with a focal distance of 0.5 m, fitted with three gratings of 300, 1800 and 3600 grooves/mm. Table \ref{Table: Grating} shows the blaze wavelengths and the equivalent full widths at half maximum (FWHM) of the slit functions (approximately a Gaussian shape) for a 2 $\mu$m wide slit measured at 355 nm for these gratings. The emission spectra from the plasma discharge have been recorded using a 1024 x 1024 pixels ICCD camera (Princeton Instruments PIMax 3) on the exit port of the spectrograph. The spectral sensitivity of the entire optical system is calibrated using the 253.65, 435.83 and 579.07 nm lines of a low pressure pencil type mercury lamp \cite{sansonetti1996wavelengths}, model ORIEL 6035. A UV-Nikkor lens with 105 mm focal length and f/4.5 is attached to the spectrometer entrance slit to collect the emitted light from the plasma discharge.

\begin{table}[h!]
\centering
\caption{Blaze wavelength and equivalent FWHM of the slit function for a 2 $\mu$m wide slit of the gratings used in this experiment.}
\label{Table: Grating}
\begin{tabular}{lccc}
\hline
\multicolumn{1}{c}{} & \multicolumn{3}{c}{Grating (grooves/mm)} \\ \cline{2-4}
                     & 300         & 1800         & 3600        \\ \hline
Blaze wavelength [nm]       & 300         & 200          & 300         \\
FWHM [nm]            & 0.58        & 0.084        & 0.037       \\ \hline
\end{tabular}
\end{table}

Synchronization between the voltage pulses signal and the ICCD camera gate was achieved by a triggering signal from the high-voltage pulse generator which was adjusted by the internal delay generator of the ICCD camera. A 200 ns time delay has been observed between the rising edge of the high-voltage pulse and the onset of the plasma discharge. This time delay comes from three sources: (1) the high-voltage pulse generator delay, (2) signal delay due to 3 meters cable length and (3) the charging of the reactor capacitance. The digital images have been recorded using a commercial 16.2 mega-pixel digital camera Nicon D5100 fitted with a micro-Nikkor lens with a 60 mm focal length and f/2.8. The high-speed images have been recoded using a FASTCAM Mini UX100 camera fitted with an Invisible vision intensifier model UVi 1850-05 and 60 mm NIKKOR micro lens. The synchronization between the image intensifier and the high-voltage pulses was achieved by a BNC-575 pulse-delay generator.


\section{Results and discussion}
\label{Results and discusion}

\subsection{Pulse characteristics and energy calculation}
\label{subsection:Electric parameters}

\begin{figure}[h]
\hspace{-1.2cm}
  \subfloat{\includegraphics[width=220pt]{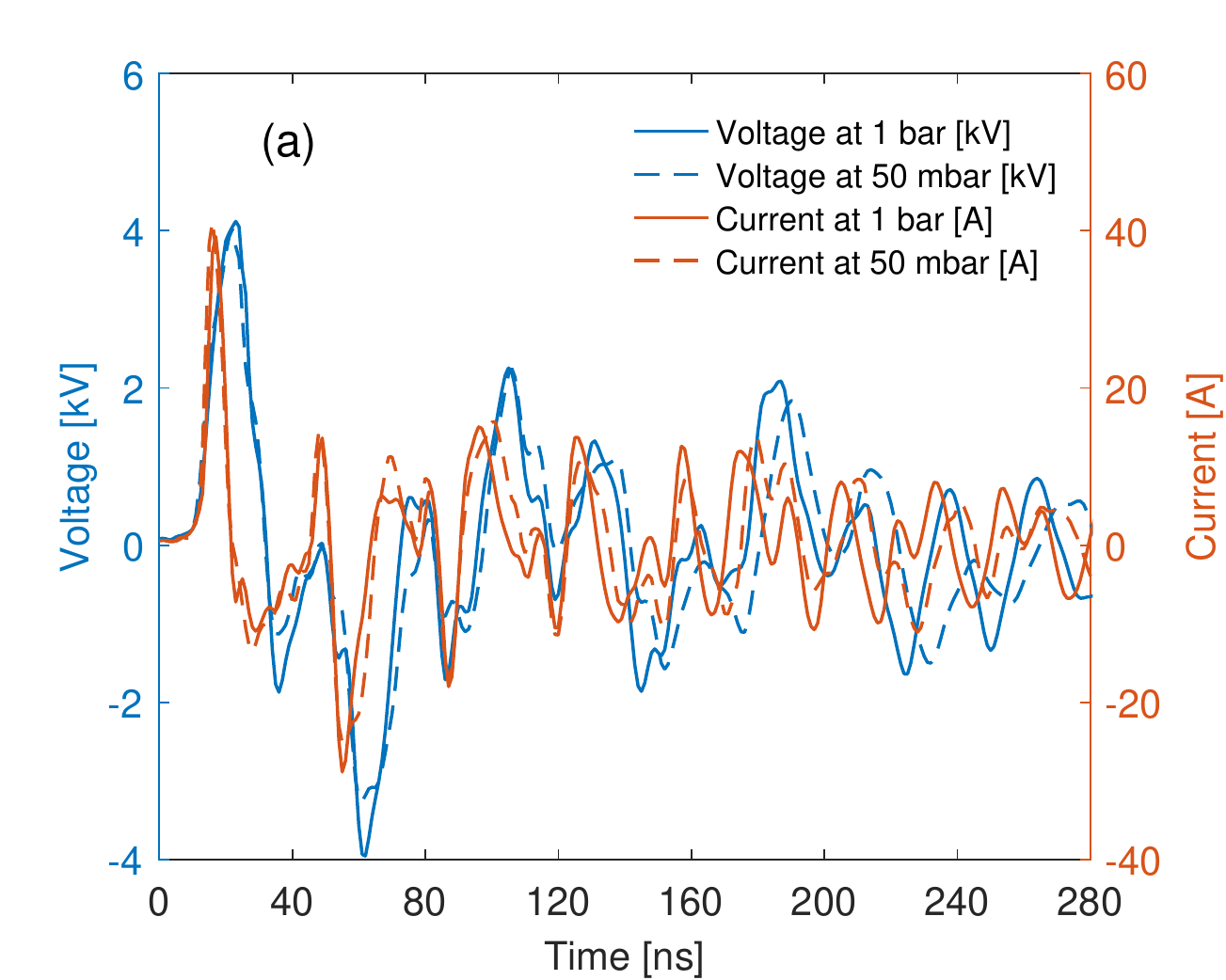}\label{fig:pulseVI}}
  \subfloat{\includegraphics[width=220pt]{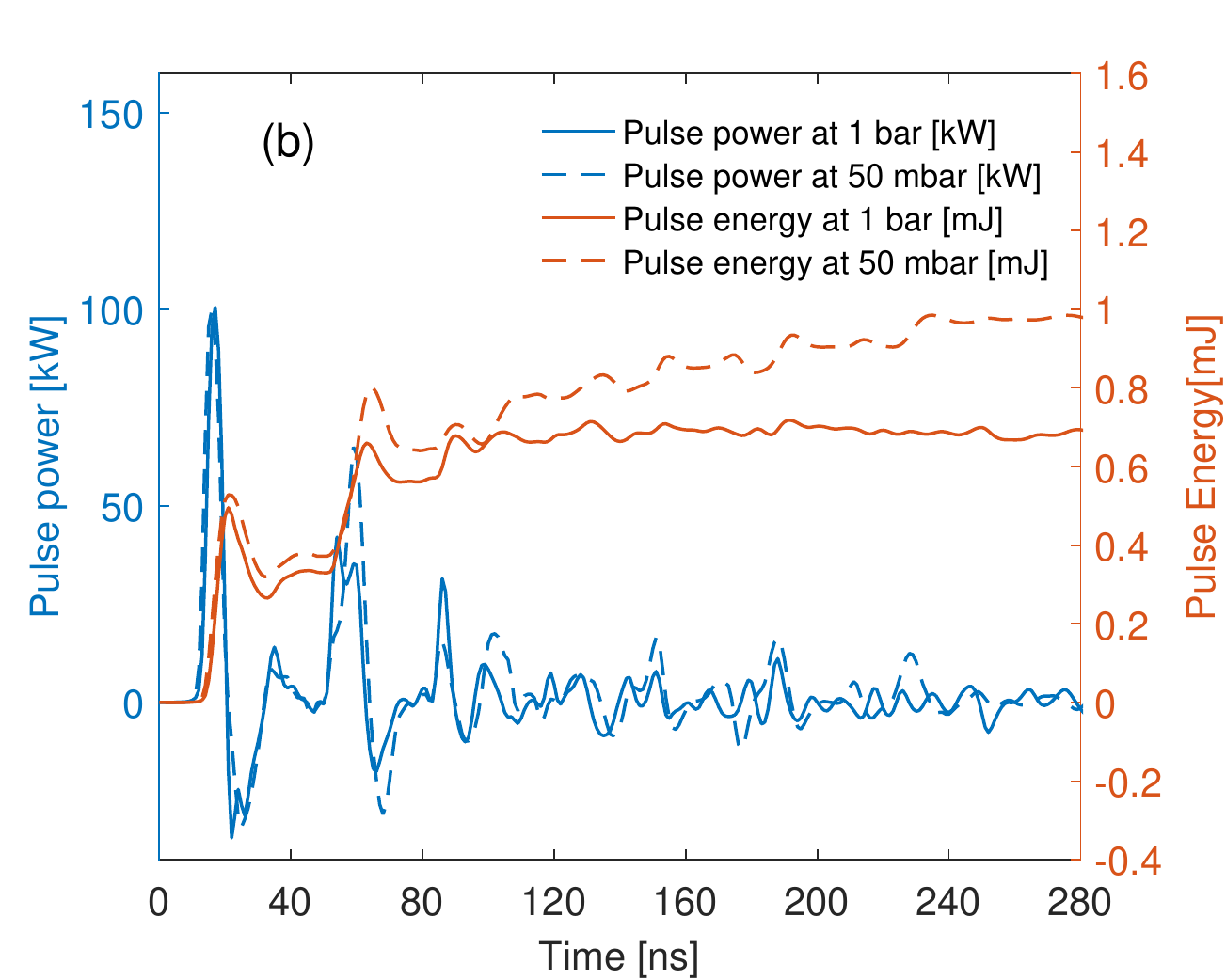}\label{figure:pulsepower}}
     \caption{\label{Fig:Power}\small (a) Temporal development the applied voltage (blue lines) and resulting current (red lines) at the plasma reactor. (b) Calculated instantaneous power (blue lines) and pulse energy (red lines). All are for stagnant air at atmospheric pressure (solid lines) and 50 mbar (dashed lines). The signals are averaged over 100 pulses.}
\end{figure}

Figure \ref{fig:pulseVI} shows the current and voltage waveforms of single pulse for a single-layer operation in stagnant air at atmospheric pressure (solid lines) and at 50 mbar (dashed lines). The signals have been averaged over 100 discharge pulses to suppress the electromagnetic noise generated by the pulse source and discharge. The voltage signal shows a 4 kV positive polarity peak followed by a few reflected waves with a rise time of 7 ns. The current signal shows the same behaviour, with a peak of 40 A, corresponding to 110 mA per hole. The full width at half maximum (FWHM) duration of the current signal is $7$ ns and it has a rise time of 4 ns.

The instantaneous pulse power $P_{pulse}$  (blue lines in Fig. \ref{figure:pulsepower}) has been obtained by multiplying the measured voltage $V$ with the current $I$. For precise calculation of the pulse power, it is important to make sure that there is no phase difference between voltage and current measurement. This was ensured by adding a temporary 75 Ohm resistance in parallel to the reactor to have a nearly resistive behaviour which makes it easy to minimize the measurement phase difference. The power has been integrated over the pulse duration to get the total energy $E_{pulse}$ deposited in the plasma discharge:

\begin{equation}	
E_{pulse} =\int_{t_1}^{t_2} I\cdot V dt.
\end{equation}

As shown in  Fig. \ref {figure:pulsepower} (red lines), the total energy per pulse is about 0.7 m${J}$/pulse at atmospheric pressure and one m${J}$/pulse at 50 mbar. Also, it has been observed that most of the power is dissipated in two peaks. The primary peak is followed by a secondary peak after about 40 ns, followed by some ripples. This behaviour has been validated by time-resolved measurement of the discharge emission, reported in section \ref{subsection:Vibrational} below.

\subsection{Discharge structure}
\label{Subsection:Discharge structure}
\begin{figure*}[ht!]
  \centering
     	\subfloat[\textit{p} = 1 bar]{
  	\includegraphics[width=100pt]{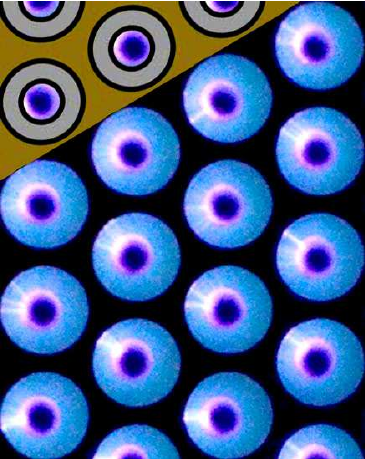}
  	\label{onebar}
  	}
\hspace{7mm}
   \subfloat[\textit{p} = 100 mbar]{
   \includegraphics[width=100pt]{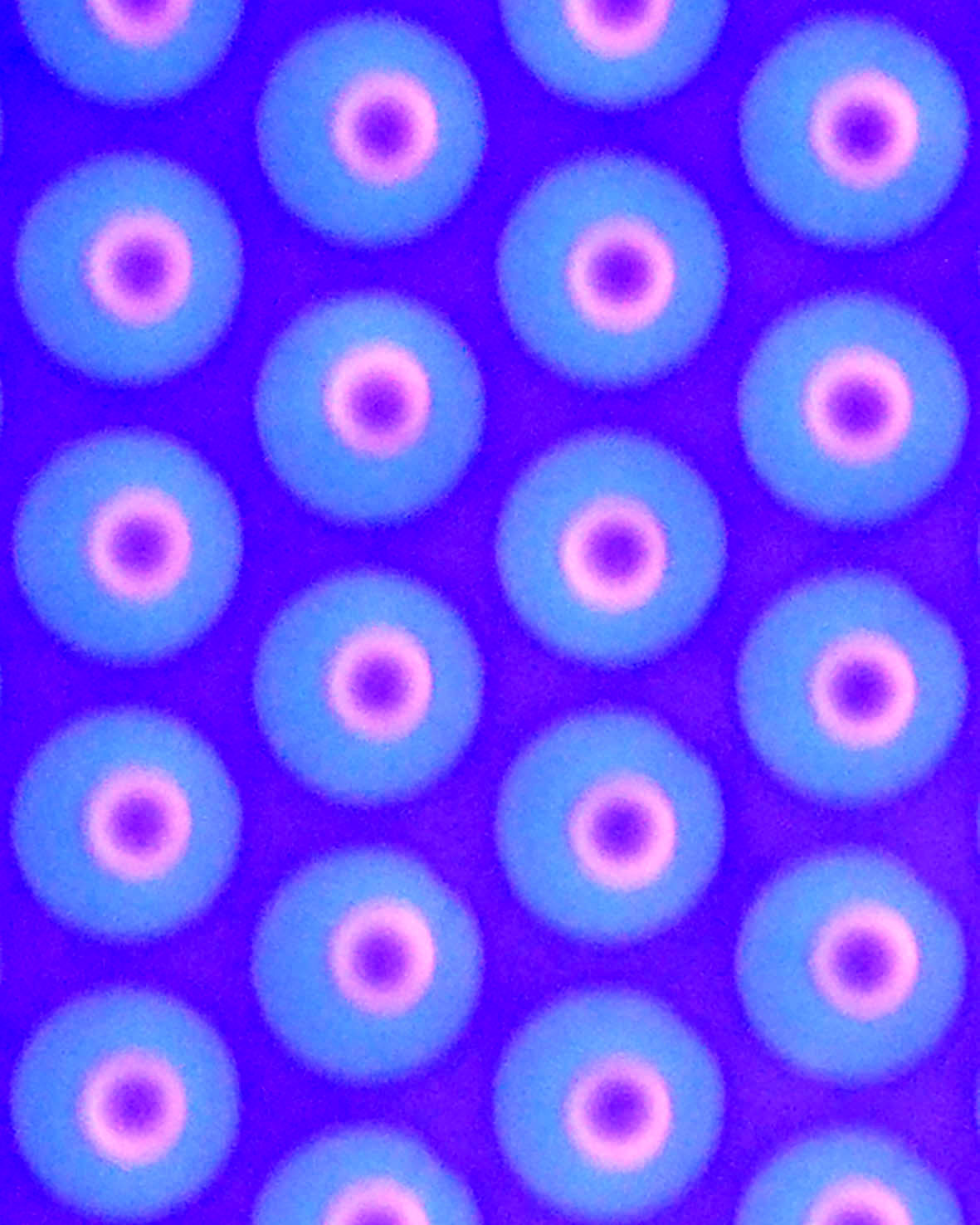}
   \label{lowbar}
   }
\hspace{7mm}
   \subfloat[\textit{p} = 50 mbar]{
   \includegraphics[width=100pt]{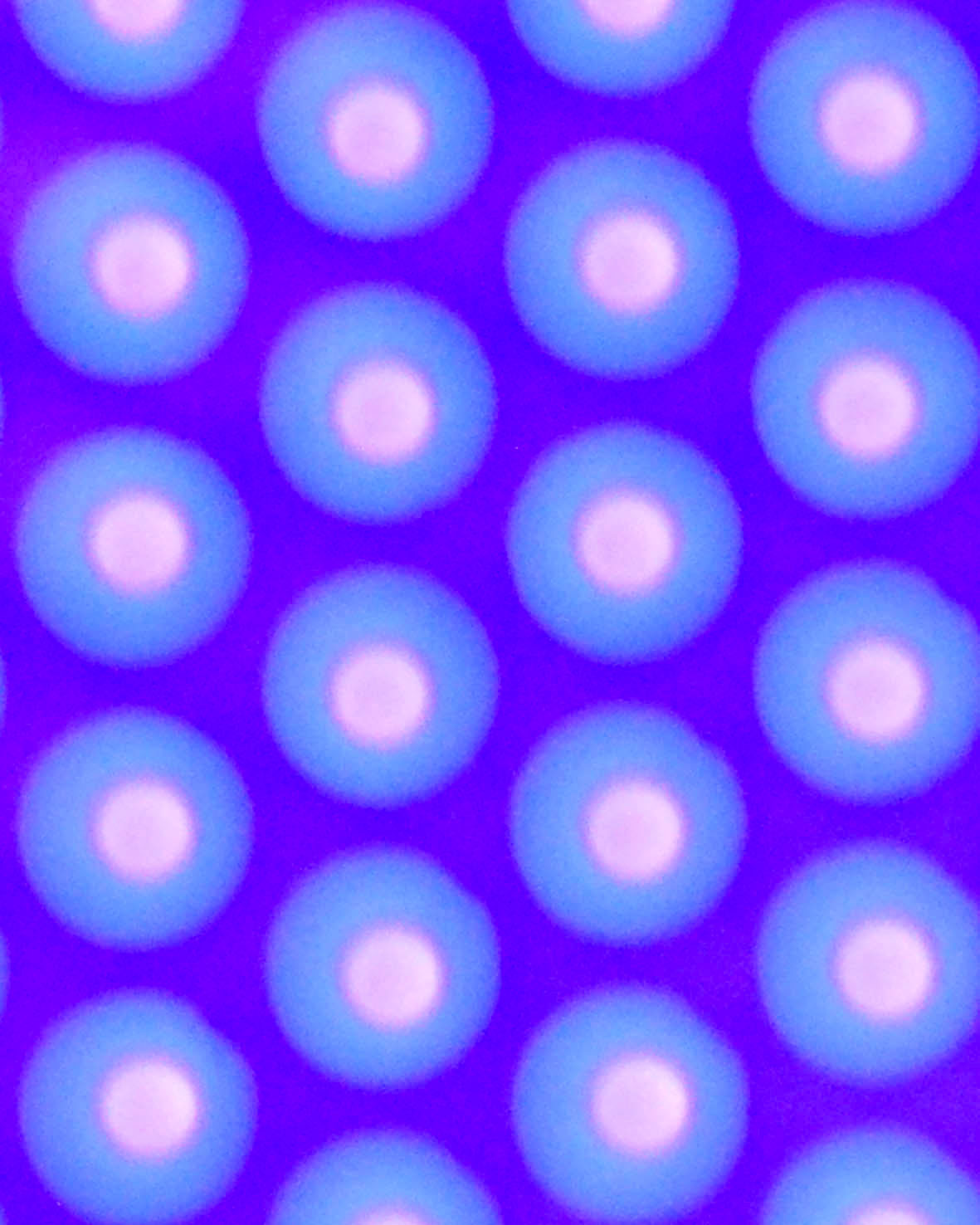}
   \label{verylow}
   }
\caption{\label{discharge structure} Natural luminosity images of plasma discharge in DBD microplasma reactor in stagnant air at voltage = 4 kV and pulse repetition frequency = 2 kHz at three different pressures. Exposure time = 50 ms. A detail of the schematic drawing from figure \ref{fig:Reactor}D has been overlayed on (a) to give a sense of scale.}
\end{figure*}

\begin{figure*}[ht!]
  \centering
     	\subfloat[\textit{p} = 1 bar]{
  	\includegraphics[width=50pt]{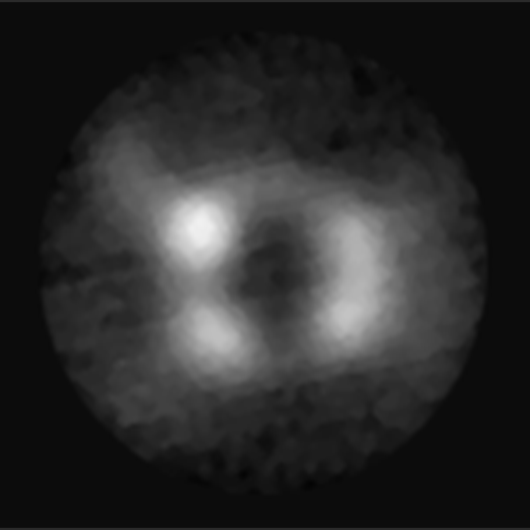}
  	\label{onebar-single}
  	}
\hspace{7mm}
   \subfloat[\textit{p} = 100 mbar]{
   \includegraphics[width=50pt]{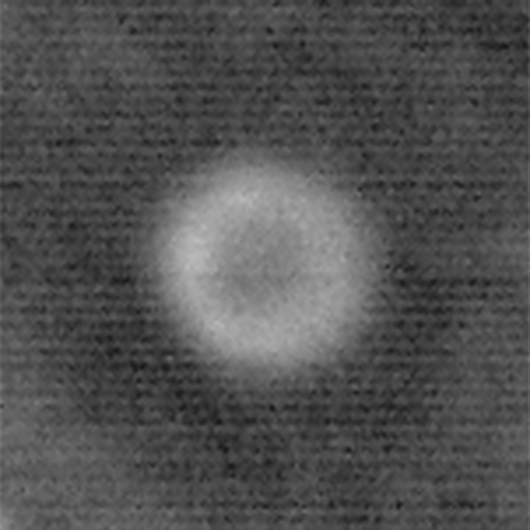}
   \label{100mbar-single}
   }
\hspace{7mm}
   \subfloat[\textit{p} = 50 mbar]{
   \includegraphics[width=50pt]{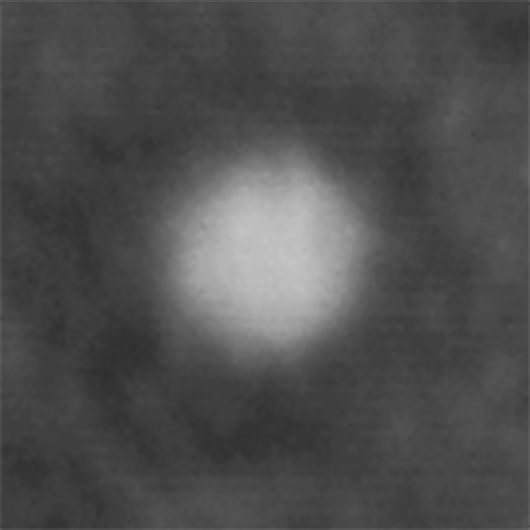}
   \label{50mbar-single}
   }
\caption{\label{single_pulse} ICCD images of a plasma discharge in a single channel during one pulse with an exposure time of 200 ns in stagnant air at voltage = 4 kV at three different pressures.}
\end{figure*}

Figure \ref{discharge structure} shows the structure of the plasma discharge inside and outside of the DBD microplasma holes at different operating pressures. The figure shows natural luminosity images of plasma reactor, looking directly into the channel from above (see Fig. \ref{fig:testrig}). The pink glow is due to the plasma, the bluish glow around it is due to the dielectric (note that the holes in the electrodes have twice the diameter of the channels). Figure \ref{single_pulse} shows short exposure ICCD images of single pulses in one channel under similar conditions.

At atmospheric pressure, Fig. \ref{onebar}, the discharge is concentrated as an annulus on the inner wall of the channel and no discharge is visible in the center of the holes. This is related to the short mean free path of the electrons due to the inelastic collisions with the neutral atoms, which narrows the discharge region. Outside the holes, a micro-filamentary discharge from the hole to the cathode ring can be observed (the white reddish lines traversing the bluish regime). The single-pulse ICCD image in figure \ref{onebar-single} shows a similar behaviour.

At a pressure of 100 mbar, Fig. \ref{lowbar}, the mean free path of the electrons increases, resulting in a wider and more uniform annular ring inside the holes. Also, note the absence of micro-filamentary discharges, also in the short exposure image (Fig. \ref{100mbar-single}). Meanwhile, the glow discharge extends beyond the confinement of the holes, giving rise to the faint pink glow over the whole field of view. By decreasing the pressure further to 50 mbar, Figs. \ref{verylow} and \ref{50mbar-single}, the plasma discharge is characterized by a high current and it uniformly fills the whole channel. Possibly, the oscillatory motion of electrons between opposite cathode fall regions, the ''pendulum effect'' \cite{schoenbach1996microhollow}, might play a role here. This would increase the number of ionization processes significantly.

 \subsection{Temperature measurements}
 \label{Subsection: Temp. measurments}
Gas temperature is a very important parameter in plasma science for two reasons. Firstly, the gas temperature has a direct effect on reactive species generation \cite{sismanoglu2010spectroscopic}. Secondly, there are many applications that need control of thermal behaviour, like wound treatment, chemical decomposition, combustion, etc. Nanosecond plasma discharges typically are of non-equilibrium nature resulting in different kinetic energy distributions for the different species (i.e. electrons, ions, neutrals). Apart from that, the different degrees of freedom (electronic, translation, vibration and rotation) are also not necessarily in equilibrium \cite{bruggeman2014gas,roupassov2009flow}. In the following sections we will focus on the rotational and vibrational temperature measurement of the plasma discharge in the DBD microplasma reactor, as well as the average gas temperature.

 \subsubsection{Rotational temperature }

 \label{Subsection: Rotational measurments}
The rotational temperature of the plasma discharge in stagnant air has been obtained by fitting the experimentally observed spectra of the $(0-2)$ band structure of the second positive system (SPS) of N$_2$ ($C ^3\Pi_u \rightarrow  B ^3\Sigma_g$) with the SPECAIR simulation tool \cite{Specair} convoluted with the measured instrumental slit function. The spectra are integrated over 20 ns following the start of the voltage pulse, using a spectrograph slit width of 2 $\mu$m. The signal has been accumulated over 300,000 pulses to increase the signal-to-noise ratio. The signal was collected from an area covering four complete holes at the center of the reactor disc. The measurements are carried out for a pulse repetition frequency of 3 kHz using the grating of 3600 grooves/mm.

\begin{figure}[h]
\hspace{-1.2cm}
  \subfloat{\includegraphics[width=220pt]{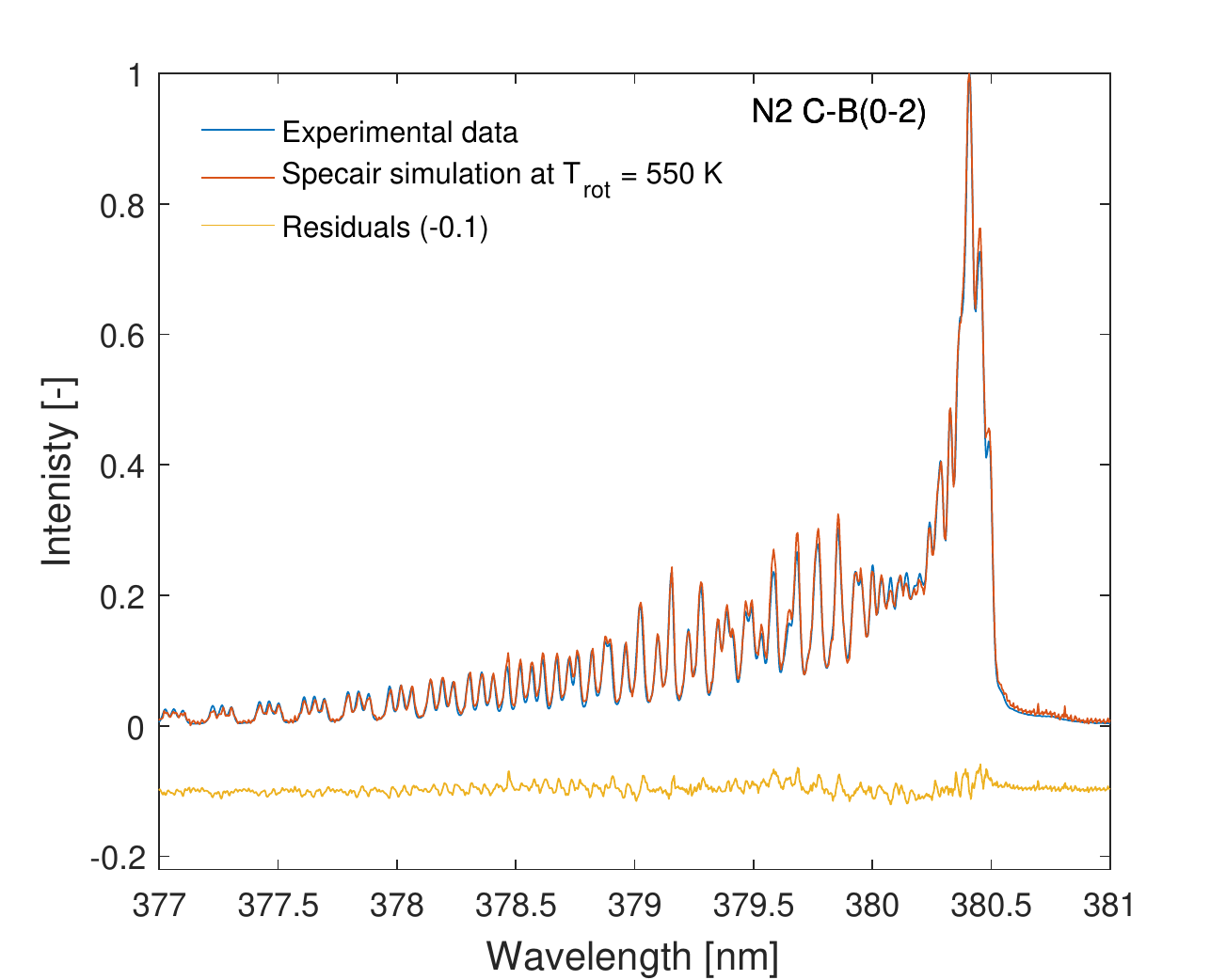}\label{R-1bar}}
  \subfloat{\includegraphics[width=220pt]{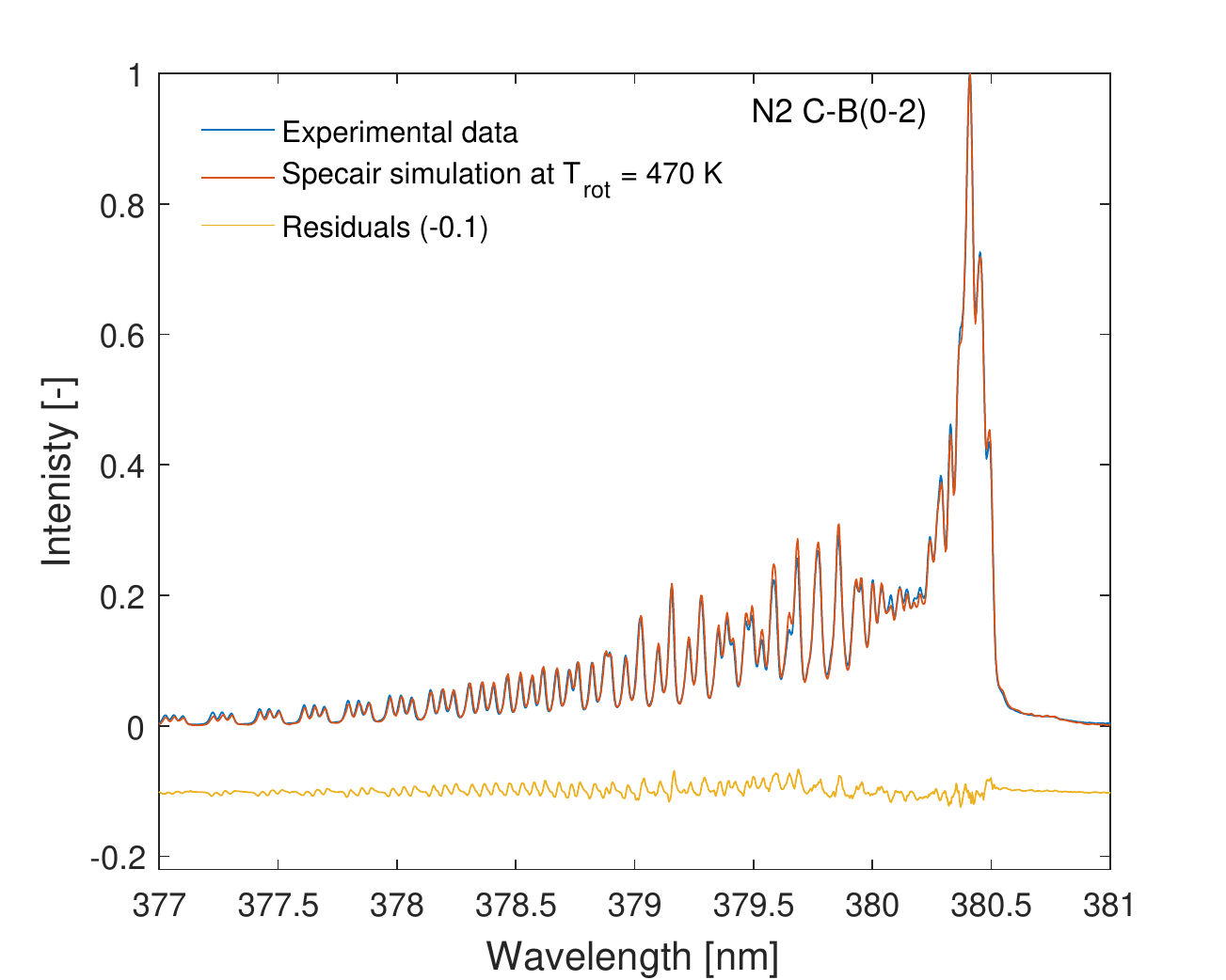}\label{R-0.1bar}}
     \caption{\label{ROT}\small Experimental (blue line) and simulated (red line) normalized spectrum of stagnant air at (left) \textit{p} = 1 bar, and (right) \textit{p} = 50 mbar. The simulated spectra are from SPECAIR with $ T_{rot}$ = 550 K and $ T_{rot}$ = 470 K respectively.}
\end{figure}

As shown in Fig. \ref{ROT}, when comparing the rotational line structure with a simulated Boltzmann rotational distribution a good fit was obtained at both atmospheric pressure and reduced pressure. This implies that the rotational distribution has thermalized (via heavy species collisions or electron collisions) in a short time (few nanoseconds). The rotational distribution of the excited states can be considered  representative of the ground state rotational temperature, which is usually close to the gas temperature \cite{fridman2008plasma}.

The rotational temperatures obtained from the SPECAIR fitting procedure were 550 K $\pm$ 30 K and 470 K $\pm$ 30 K for pressures of \textit{p} = 1 bar and \textit{p} = 50 mbar, respectively. According to Fig. \ref{figure:pulsepower}, the consumed power in the plasma at reduced pressure is higher than at atmospheric pressure. This would indicate that the temperature is higher at reduced pressure. However, the lower collision frequency, apparently, ensures that the energy stays more confined to other degrees of freedom like vibrational energy as will be shown in section \ref{subsection:Vibrational}.

\subsubsection{Vibrational temperature}
\label{subsection:Vibrational}
The vibrational temperature describes the population of the vibrational states of molecular species, if this follows a Boltzmann distribution. Unlike the rotational excitations, vibrational excitations are mainly driven by electron collisions and, therefore, the vibrational population provides some information about the electron energy.

In this study, the vibrational temperature has been obtained by fitting the observed structure of the (0-2), (1-3) and (2-4) spectral bands of the SPS of N$_2$ to SPECAIR simulations. By using a grating of 1800 grooves/mm and integrating the signal over 100,000 pulses.

\begin{figure}[h]
\hspace{-1.2cm}
  \subfloat{\includegraphics[width=220pt]{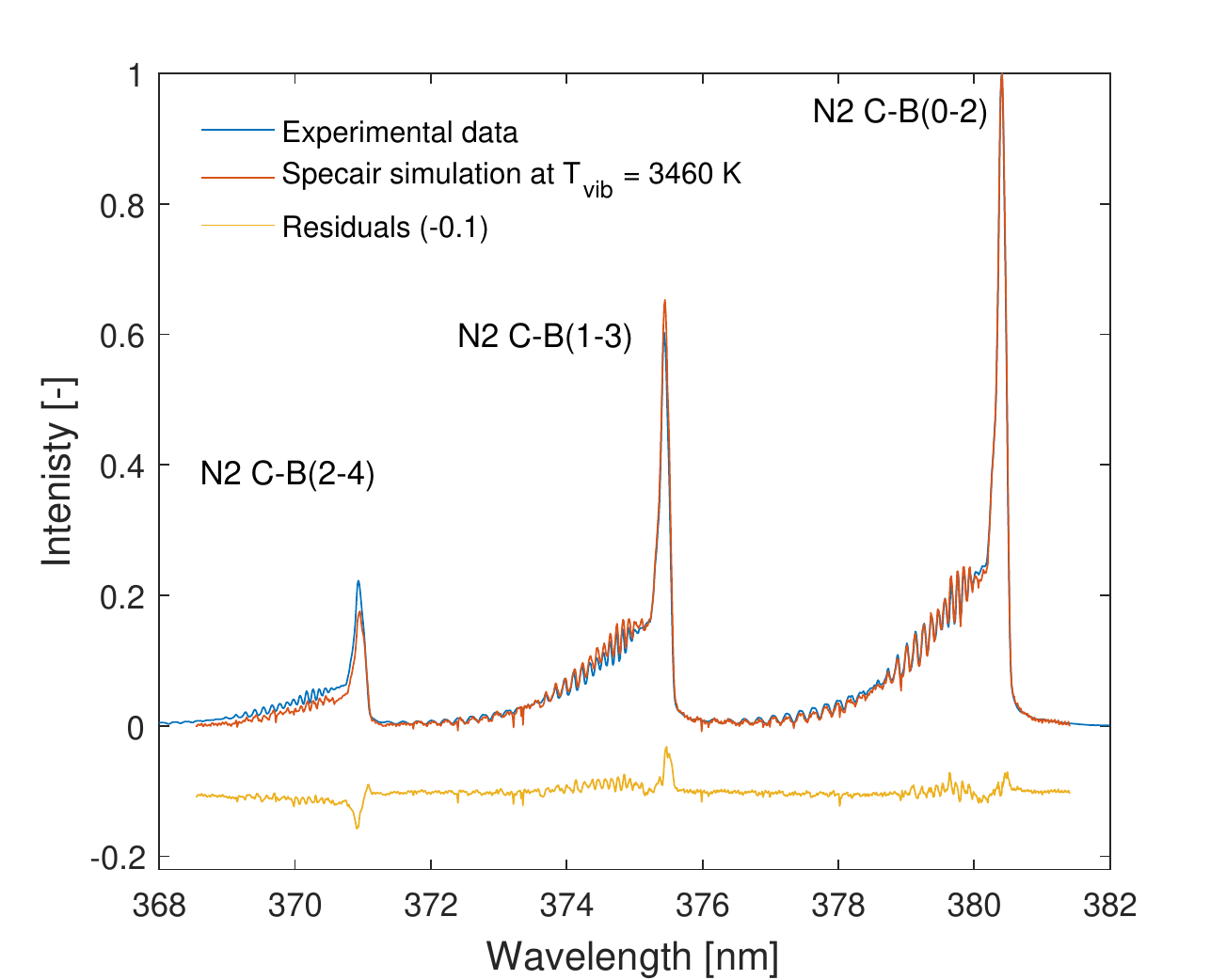}\label{V-1bar}}
   \subfloat{\includegraphics[width=220pt]{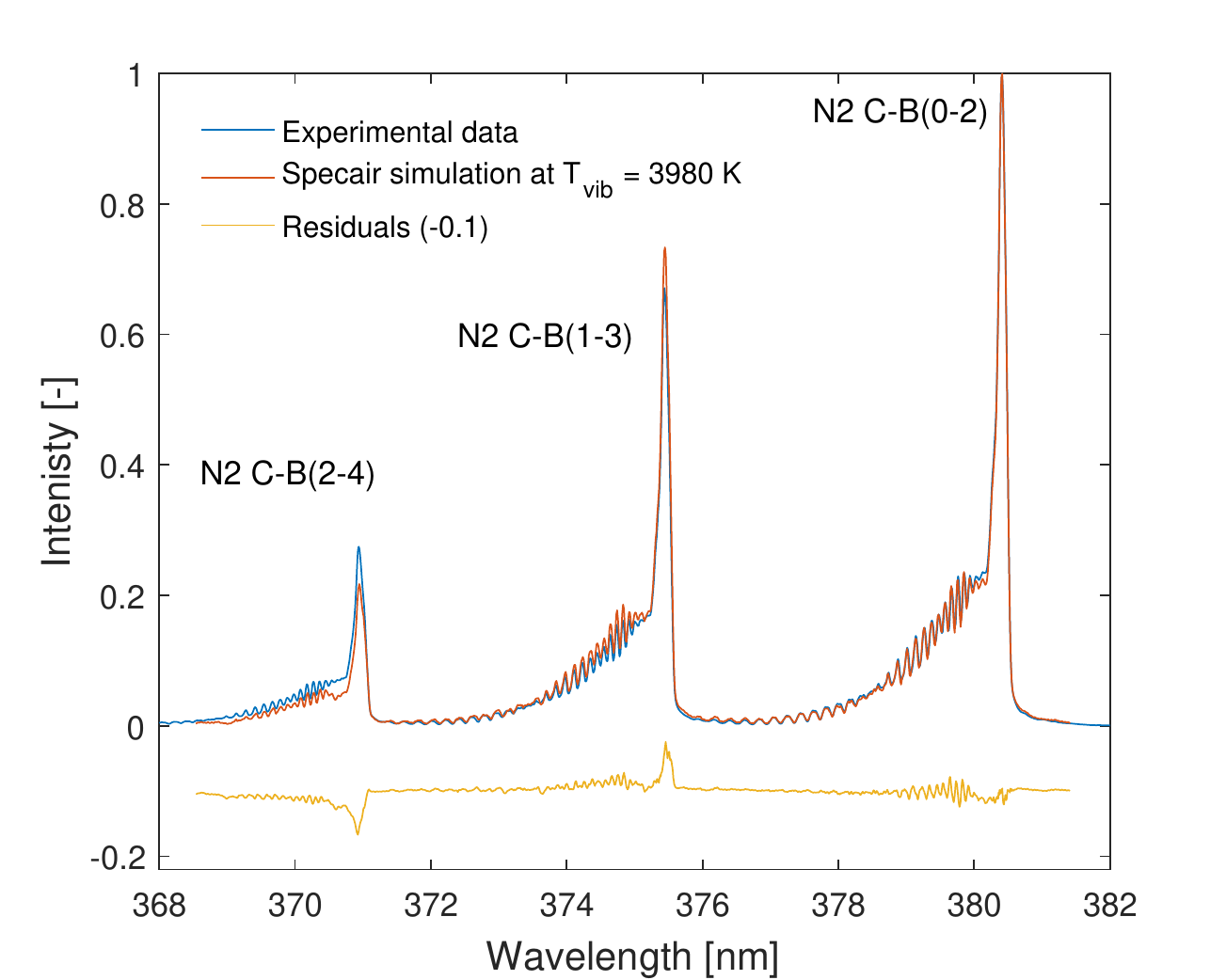}\label{V-0.1bar}}

     \caption{\label{VIB}\small Experimental (blue line) and simulated (red line) normalized spectra of stagnant air at (left) \textit{p} = 1 bar, and (right) \textit{p} = 50 mbar. The simulated spectra are from SPECAIR with $ T_{vib}$ = 3460 K and $ T_{vib}$ = 3980 K respectively.}
\end{figure}

Here, the fit is not as good as for the rotational temperature, leaving a higher uncertainty in the vibrational temperature calculations. This fitting error may be caused by errors in the intensity calibration of the spectrometer or by the non-Boltzmann nature of the discharge. Therefore, the estimated vibrational temperature here is the temperature corresponding to the minimum residual error for these three bands considered. As shown in Fig. \ref{VIB}, estimated vibrational temperatures of $3460 \pm 100$ K at atmospheric pressure and $3980 \pm 100$ K at \textit{p} = 50 mbar can be extracted. Obviously, the vibrational temperature is considerably higher than the rotational and, presumably, the translational temperatures. The reduction of the vibrational temperature with increasing operating pressure can be attributed to two factors. On the one hand, there is enhanced collisional relaxation of the vibrationally excited levels as the collisional frequency increases. On the other hand, less energy may be put into vibration due to the reduction in the electron energy with pressure. This finding will be discussed in more detail later, in the context of estimating the reduced electric field strength at different operating pressures in section \ref{E/N}.

\begin{figure}[h!]
 	\centering
 	\includegraphics[width=280pt]{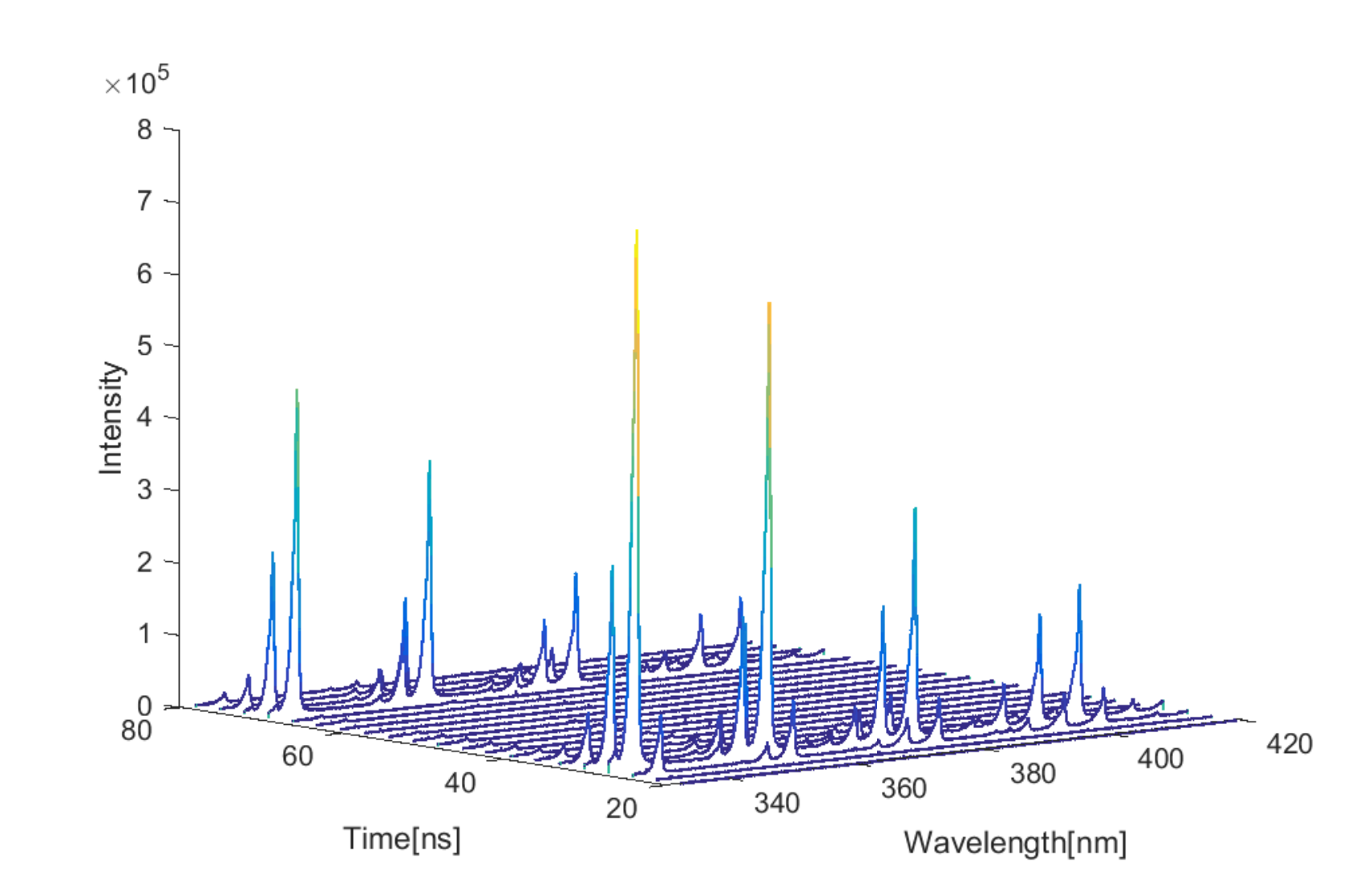}
	\caption{\label{Spectra3D}\small Temporal development of emission spectra of the plasma discharge at atmospheric pressure in air, with an integration window of 3 ns using a 300 grooves/mm grating. }
	
\end{figure}

\begin{figure}[h]
 	\centering
	\includegraphics[width=250pt]{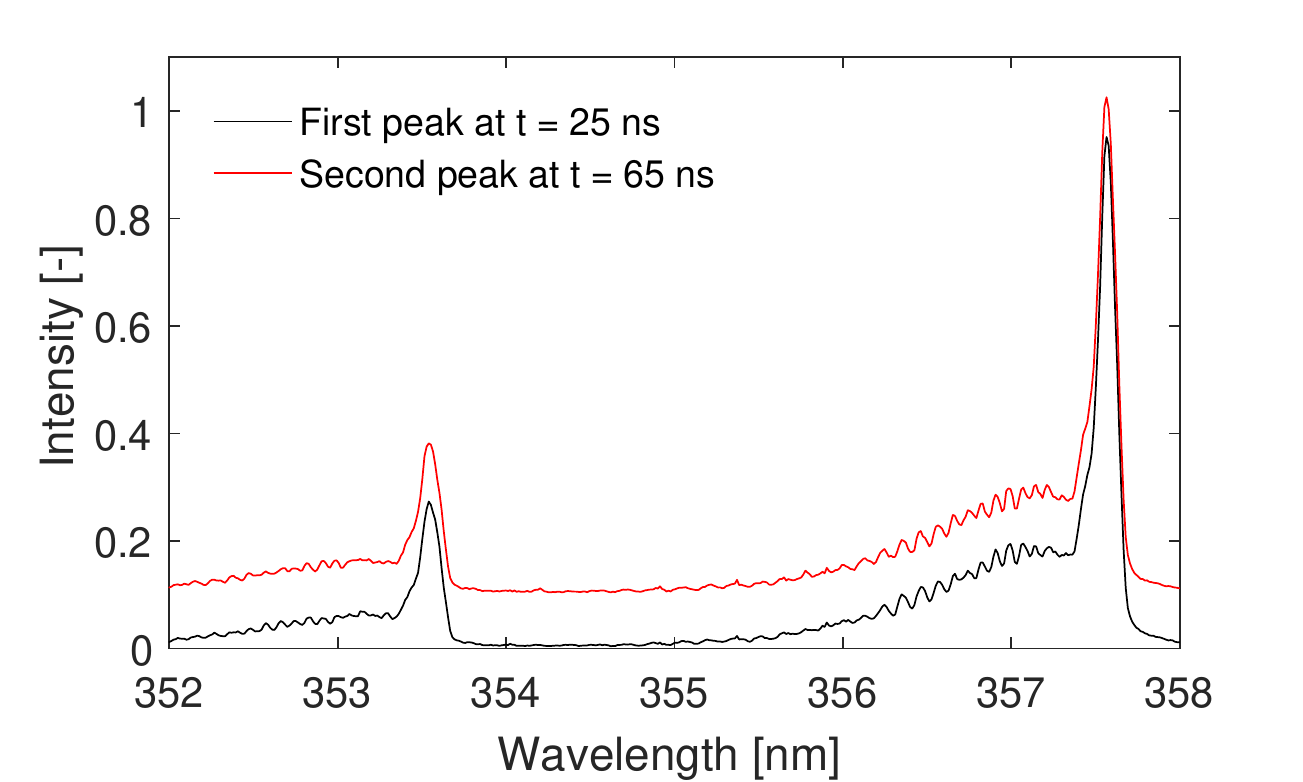}
	\caption{\small A comparison of the normalized spectra for the first peak (black line) and the second peak (red line) during one pulse. The signals are integrated over 100,000 pulses for stagnant air at atmospheric pressure. The second peak line has been shifted upward by 0.1 for better visualization.}
	\label{Fig:pulsewave}
\end{figure}

In order to illustrate the discharge behaviour during the pulse duration, the time evolution of spectral intensity between 330 nm and 420 nm over a time span of 80 ns is shown in Fig. \ref{Spectra3D}. Each spectrum in this figure is integrated over a temporal window of 3 ns, using a 300 grooves/mm grating. As shown, there are two main peaks in the spectra, separated by 40 ns. This behaviour agrees very well with the power measurements from Fig. \ref{figure:pulsepower}, as it also shows two principle power peaks with roughly the same time separation.

It is worth to mention that no vibrational or rotational temperature differences were observed between the first peak at t = 25 ns and the second peak at 65 ns, as can be seen in Fig. \ref{Fig:pulsewave}. These spectra have been recorded in stagnant air at atmospheric pressure using the 1800 grooves/mm grating centered at 355 nm to cover the (0-1) and (1-2) bands of the SPS of N$_2$. About $30$ percent reduction in emission intensity has been observed between the first and the second peak.


\subsubsection{Average gas temperature}
\begin{figure}[h!]
\centering
\includegraphics[width=250pt]{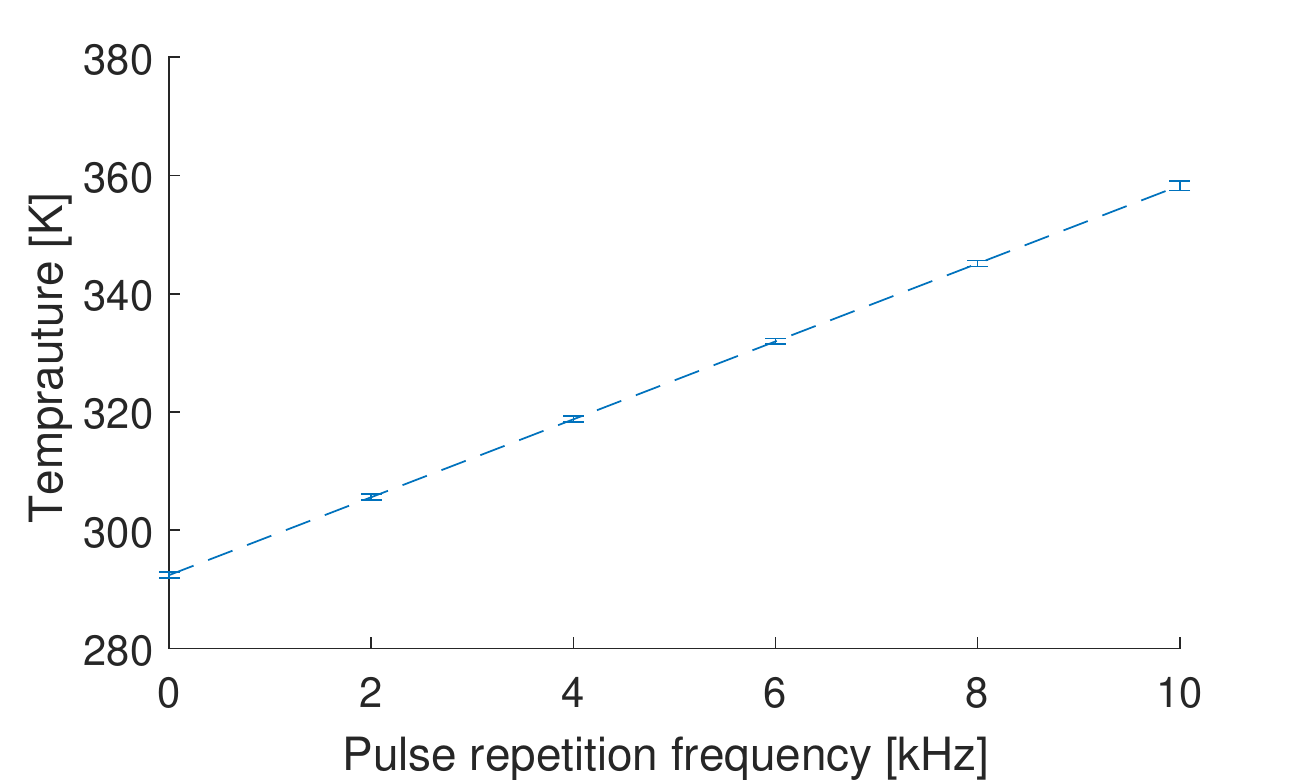}
\caption{\small Measured gas temperature as function of pulse repetition frequency for air with flow rate of 3.4 l/min at atmospheric pressure and V = 4 kV. The dashed line indicates a linear fit and the error bars represent the standard deviation in the measurements.}
\label{Air_temp}
\end{figure}

To measure the average gas temperature downstream of the reactor, we have used a thermocouple type K (model Testo 0613-1912) with a 5 mm probe diameter and a reaction time of 35 seconds to damp noise. The probe junction is located 5 mm downstream from the central hole of the reactor. Fig. \ref{Air_temp} shows the air temperature as a function of pulse repetition frequency for an air flow rate of 3.4 l/min at atmospheric pressure.

As shown in the figure, there is a linear relationship between pulse repetition frequency and average gas temperature, with a slope of 6.5 K/kHz. From this relationship, we can extract an average value of the thermal power generated by the plasma discharge, $Q_{plasma} = \dot{m} \cdot c_p \cdot \Delta{T}$, where $\dot{m}$ is the mass flow rate, $c_p$ is the specific heat of air at constant pressure and $\Delta{T}$ is the difference between inflow and outflow temperatures. This gives an average value of 0.45 W for 4 kV pulses at a repetition frequency of 1 kHz, or in other words, the energy spent to gas heating is 0.45 mJ/pulse for all channels combined.

This value is 65 \% of the pulse energy calculated from the voltage and current measurements at atmospheric pressure in section \ref{subsection:Electric parameters}. The rest of the consumed power is spent on other degrees of freedom such as dissociation of the molecular gas components (like the dissociation of the molecular oxygen in the quenching reaction of N2(B) \cite{rusterholtz2013ultrafast}), heating of the dielectric material and conduction to the metal body. In the case of reduced pressure, it is expected that the fraction of the power which is dissipated into gas heating will be lower, and more power will go to the chemical processes due to the higher electron energy and hence ionization efficiency.


\subsection{Effective reduced electric field strength $\EN$}
\label{E/N}
Determination of the actual reduced electric field (\textit{E/N}) by direct methods is relatively complicated for pulsed microdischarges because both the electric field (\textit{E}) and the concentration of neutral particles (\textit{N}) vary strongly in time and space. However, Paris et al. \cite{Paris} introduced an indirect method by using the intensity ratio of nitrogen spectral bands. 
For this the (0-0) transition band of the first negative system of N$^+_2$ (FNS; $B ^2\Sigma^+_u \rightarrow  X ^2\Sigma^+_g$) at 391.4 nm and the SPS (2-5) band at 394.3 nm are used. The following relation between the intensity ratio $R_{391/394}$ and the reduced electric field has been proposed for low-temperature plasmas in air at a pressure range of $0.3 - 100$ kPa with a maximum deviation of 12\%:

\begin{equation}
\label{EQ:line_ratio}
\mathrm{\textit{R}_{391/394} = 46 \ exp\Bigg[-89 \bigg(\frac{\textit{E}}{\textit{N}}\bigg)^{-0.5}\Bigg]}.
\end{equation}

In this section, this method has been implemented to determine the effective reduced electric field strength $\EN$ during 100 ns after the onset of the pulse and averaged across the entire channel diameter.

\begin{figure}[h]
 	\centering
 	\includegraphics[width=250pt]{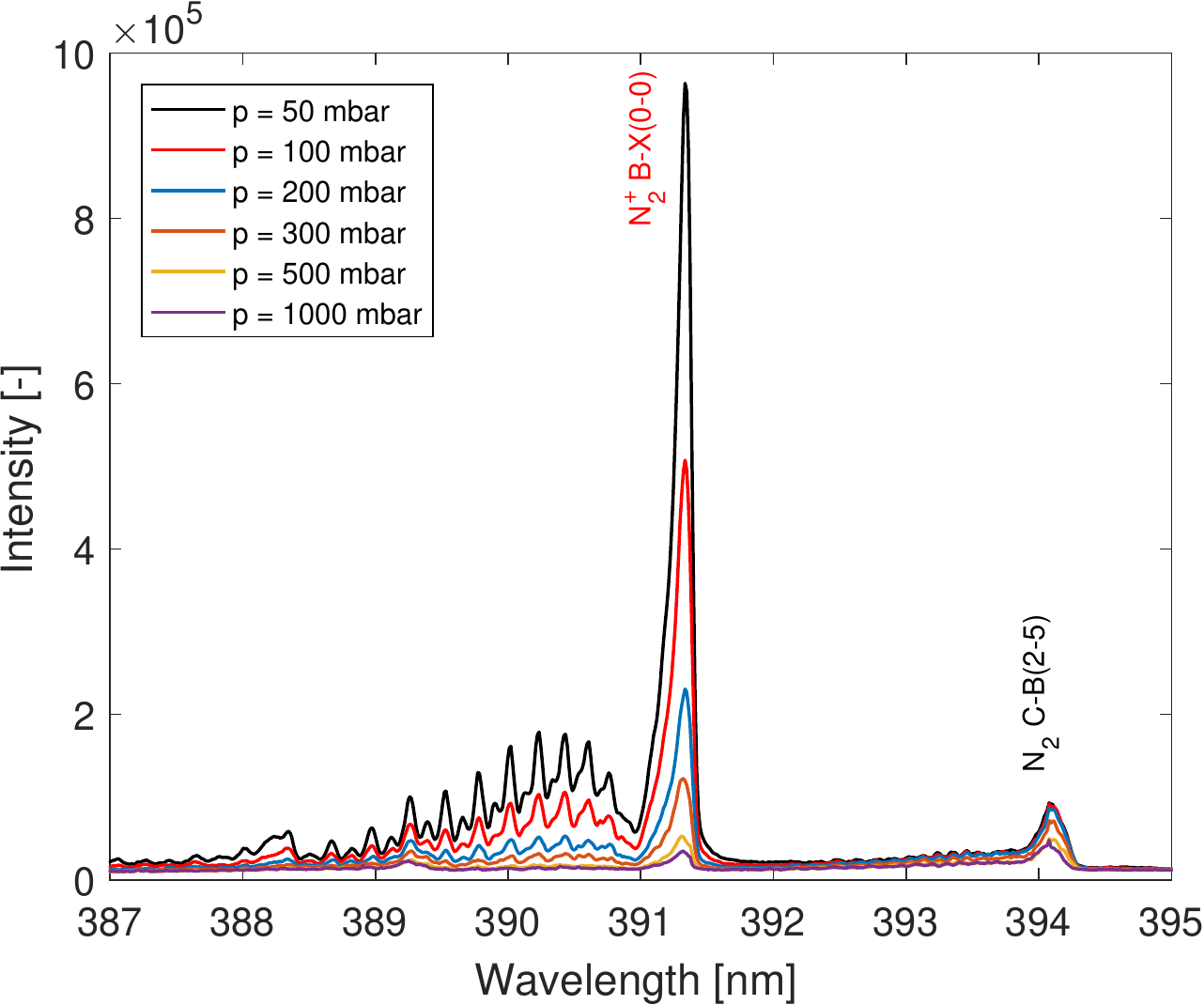}
	\caption{\small Emission spectra showing the 391.4 nm and 394.3 nm nitrogen bands at different operating pressures using the 1800 grooves/mm grating, 4 kV pulses at 3 kHz repetition rate in stagnant air with an integration window of 100 ns.}
	\label{Fig:ENspectra}
\end{figure}

\begin{figure}[H]
 	\centering
 	\includegraphics[width=250pt]{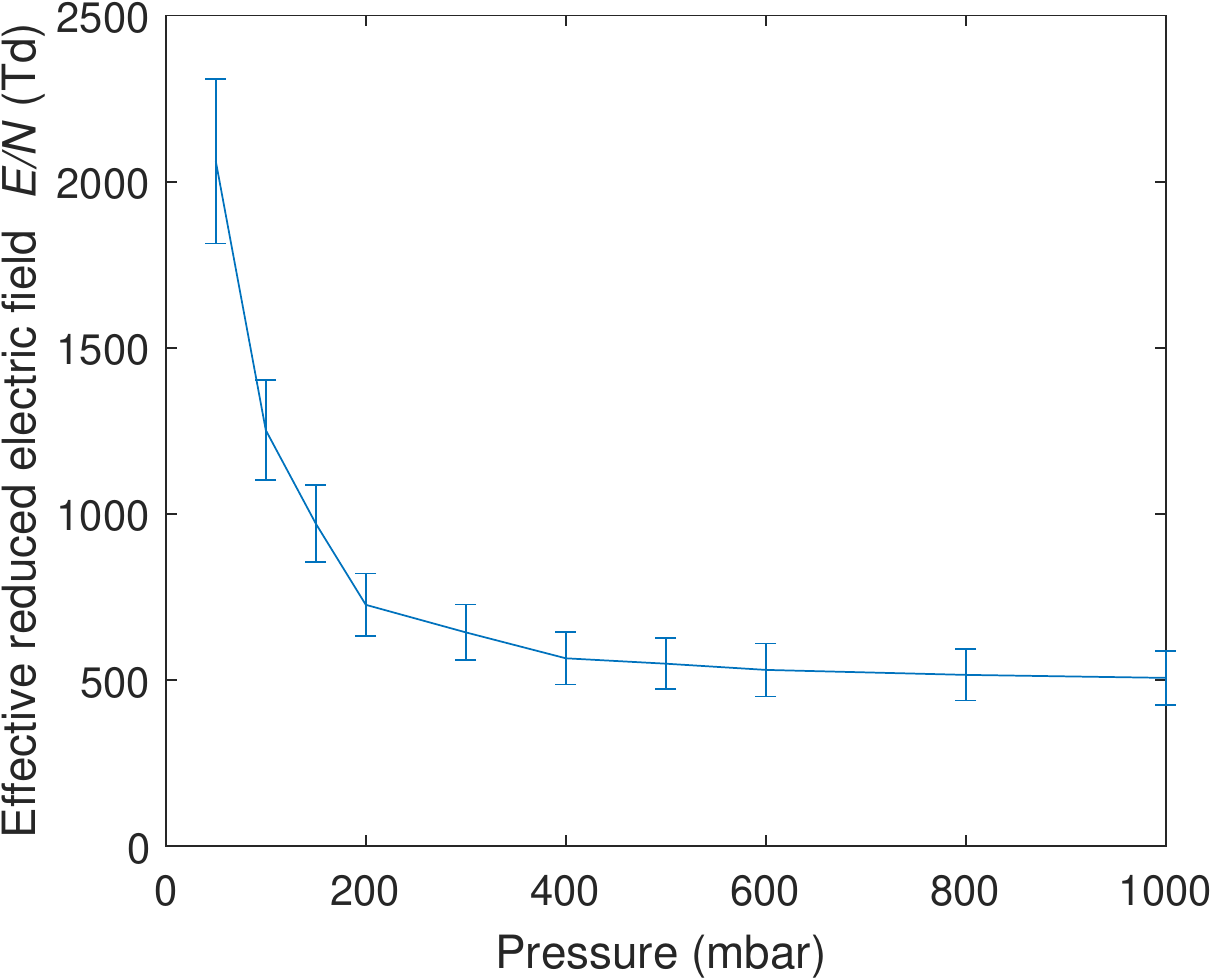}
	\caption{\small  Effective reduced electric field $\EN$ as a function of the operating pressure calculated from the intensity ratio $R_{391/394}$. Plasma conditions are given in Fig. \ref{Fig:ENspectra}. The error is estimated by the maximum uncertainty of the method (12 $\%$) in addition to the sensitivity error in the peaks determination.}
	\label{Fig:EN}
\end{figure}

Fig. \ref{Fig:ENspectra} shows the intensities of the N$^+_{2} (B-X) $ (0-0) and N$_{2} (C-B)$ (2-5) bands of nitrogen over the pressure range of 1 bar to 50 mbar. Fig. \ref{Fig:EN} shows the effective $\textit{(E/N)}$ obtained by equation \ref{EQ:line_ratio}, as a function of operating pressure. In this figure, we can see that at operating pressures between 1 bar and 400 mbar, the effective reduced electric field $\EN$ increases gradually from $500 \pm 100$ Td $600 \pm 100$ Td. By reducing the pressure down to 50 mbar, the $\EN$ increases steeply up to $2050 \pm 250$ Td. This is due to the reduction in gas density which leads to an increase in electric field as discussed in section \ref{subsection:Electric parameters}. This will be accompanied by an increase in average electron energy.

\subsection{Effect of flow rate on discharge emission}

To study the effect of the flow velocity inside the holes on the discharge intensity in the DBD microplasma reactor, a set of experiments has been done at different air flow velocities at atmospheric pressure with 4 kV pulses at 3 kHz repetition rate. This study has been carried out for mean air flow velocities ranging from 0 to 7 m/s inside the channels, corresponding to average flow velocities from 0 to 1 m/s above the reactor.

\begin{figure}[h]
 \centering
   \begin{tikzpicture}
     \node at (1,1) {\includegraphics[width=250pt]{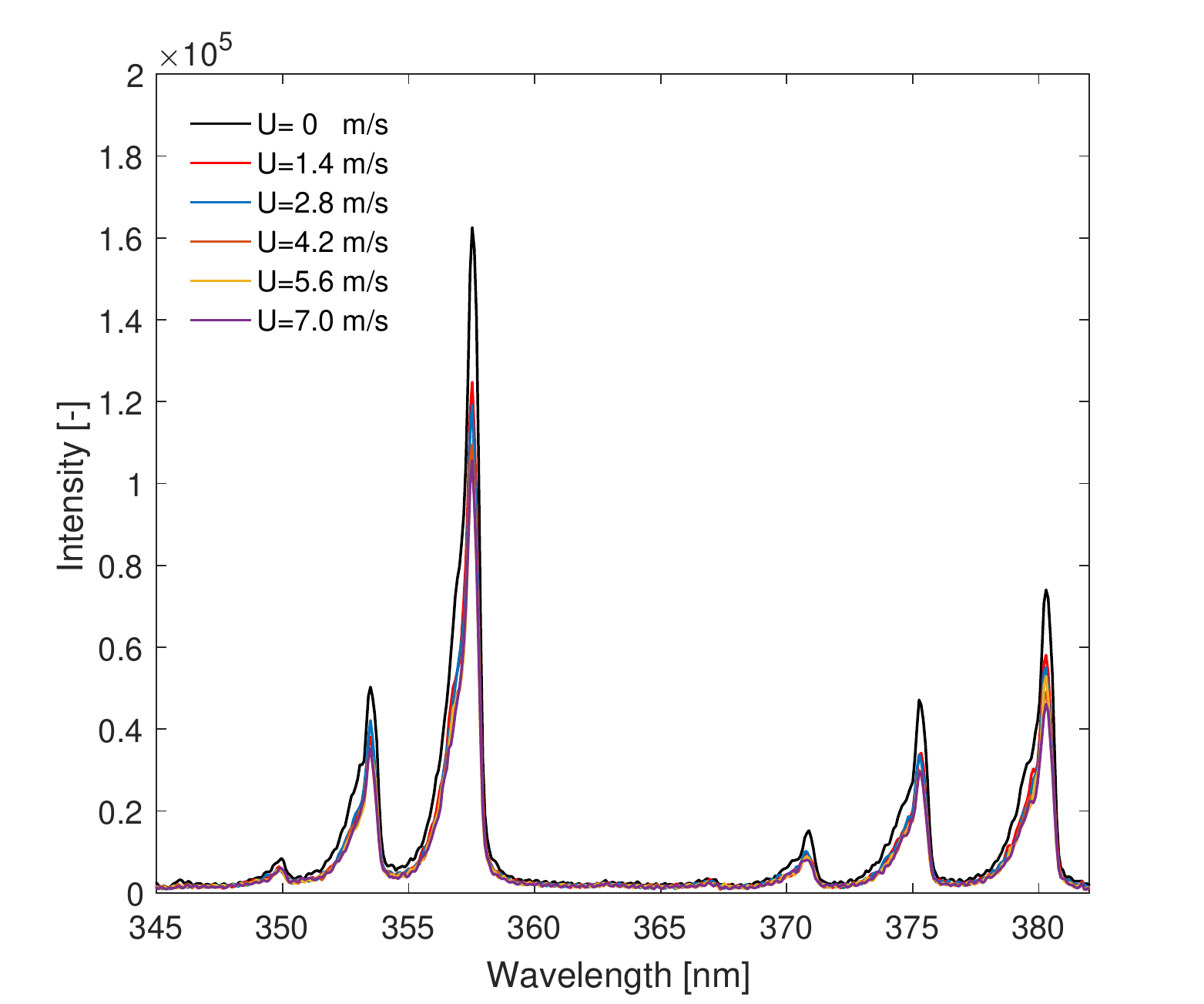}};
     \node at (2.5,1.9) {\includegraphics[scale=0.22]{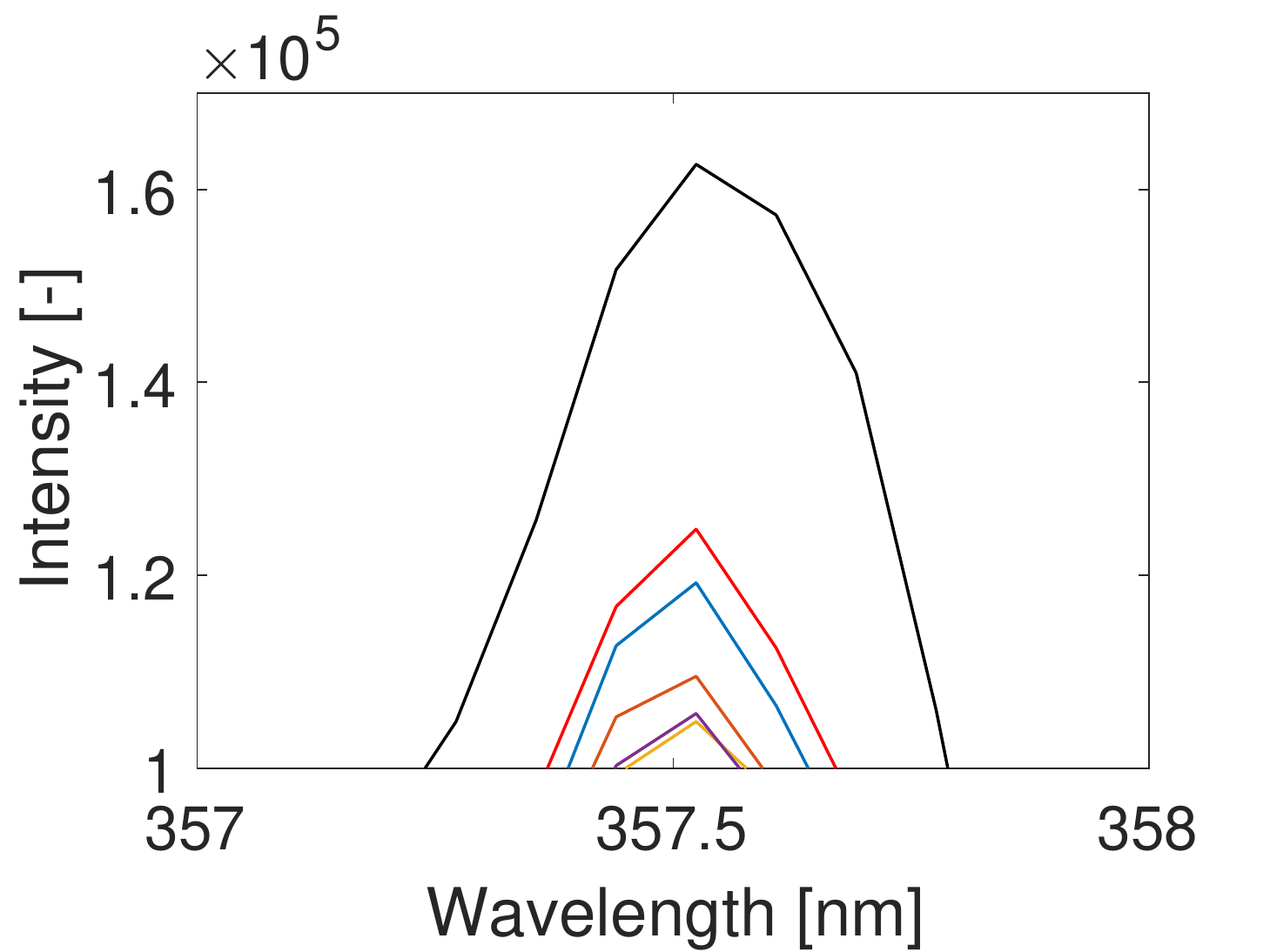}};
     \draw [red,dashed] (-0.15,1) rectangle (0.25, 3.5) ;
  \end{tikzpicture}
\caption{\small Non-normalized emission spectra in air at atmospheric pressure for different flow velocities for 4 kV pulses at 3 kHz repetition rate. The inset shows a magnification of the largest peaks (dashed).}
\label{Fig:flow}
\end{figure}

The residence time of a volume of gas passing through the discharge region in a channel (length of 0.36 mm for single-layer operation) ranges from 50 to 250 $\mu$s for air velocities between 7 and 1.4 m/s. Obviously, the minimum residence time is much longer than the characteristic time scale of the plasma discharge, which is on a tens of nanosecond time scale. However, these residence times are slightly shorter than the inter-pulse delay at a pulse repetition frequency of 3 kHz. This can explain the reduction in emission intensity at higher flow velocities, as shown in Fig. \ref{Fig:flow}. The figure shows a roughly 25\% reduction in emission intensity when increasing the flow from 0 to 1.4 m/s and a smaller, more gradual, reduction when increasing the flow further up to 7 m/s. In stagnant air, all high-voltage pulses ionize the same volume of gas in the discharge region. This increases the ionization level before the new breakdown occurs which leads to higher emission intensity \cite{hoft2016impact,Sander} while at higher velocities most of the gap is filled with fresh gas and less pre-ionization will exist.

\subsection{Reactor lifetime assessment}
\label{sec.lifetime}
One of the serious problems of microplasma devices is their often short lifetime due to surface and internal damages. This is due to the frequent loss of plasma confinement due to vaporization and ablation, which cause significant surface erosion and structural failure \cite{lifetime,hassanein1998performance}. Reactor lifetime is very sensitive to many parameters, e.g. geometry, material of the dielectric and power source.

\begin{figure}[h]
  \centering
  \includegraphics[width=250pt]{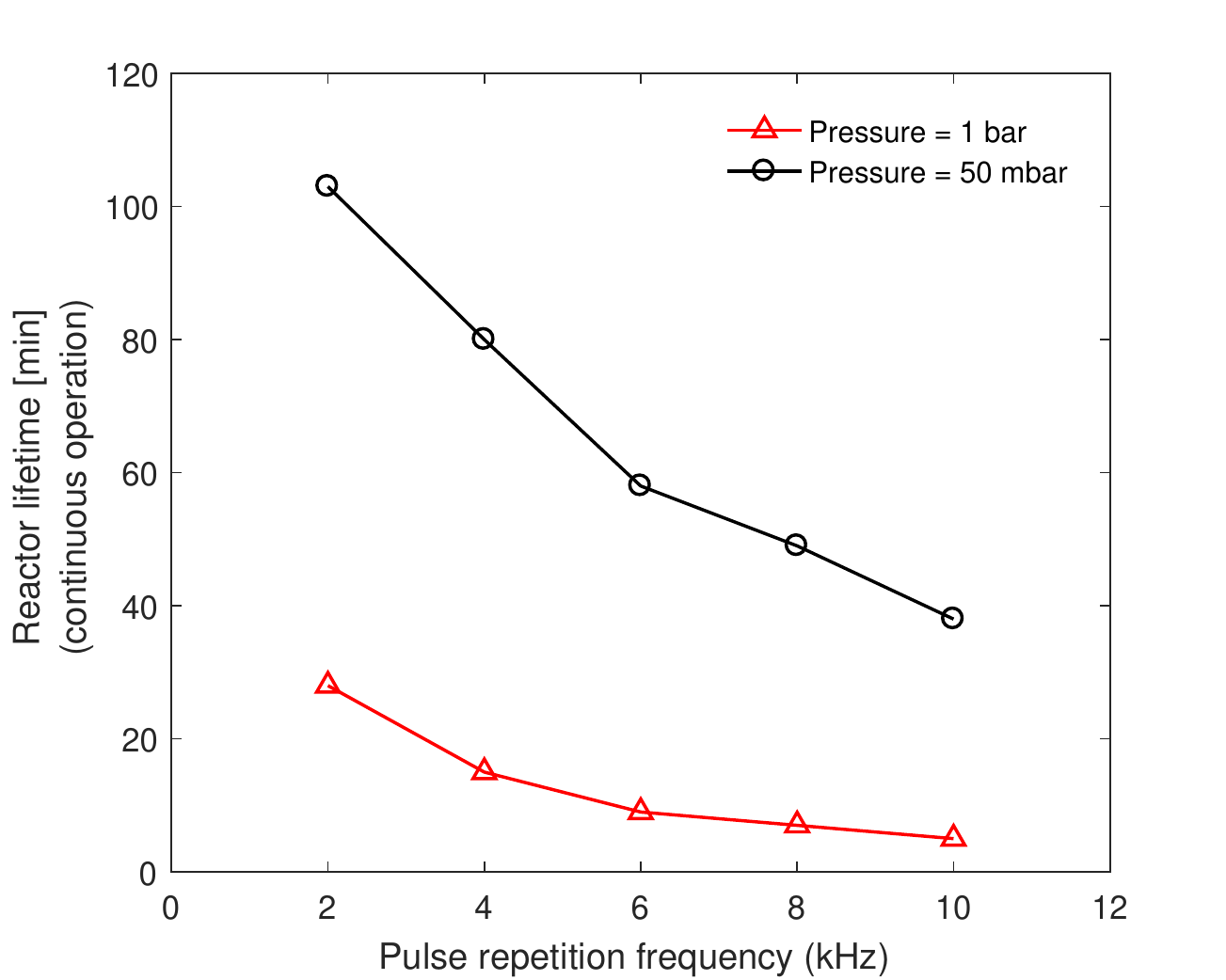}
  \caption{\label{fig:lifetime} Reactor lifetime in continuous operation mode as function of pulse repetition frequency at atmospheric pressure and 50 mbar.}
\end{figure}

Fig. \ref{fig:lifetime} shows the dependence of the reactor lifetime on the pulse repetition frequency at atmospheric pressure and 50 mbar. At a frequency of 2 kHz and atmospheric pressure, the reactor lifetime was about 24 minutes in continuous operation mode. While at 100 mbar, it was about 103 minutes. This shows an enhancement in the reactor lifetime of about four times by lowering the operating pressure by a factor of 20. This enhancement increases to a factor of about ten at a pulse repetition rate of 10 kHz, even though the lifetime itself decreases in both cases.

Two factors may contribute to the observed enhancement. Firstly, at low pressure, the ion impact frequency is low, which slows down the dielectric wear down process. Secondly, at low pressure the gas temperature is lower as was shown in section \ref{Subsection: Rotational measurments}. Both reasons make the reactor more prone to failure at atmospheric pressure due to thermal stress on the internal structure of the dielectric material. Other factors that may affect the lifetime of the reactor and need more investigation are voltage amplitude and duration, geometry, reactor material and operating gas.


\section{Summary and conclusion}
In this work, a new DBD microplasma flow reactor driven by nanosecond high-voltage pulses has been presented. By this design we were able to sustain a non-equilibrium plasma discharge in a pattern of 400 $\mu$m diameter channels at pressures up to 1 bar. The DBD microplasma flow reactor is a promising device for a wide range of applications that require an efficient interaction between a non-thermal plasma discharge and a gas flow. Some of these applications are pollutant control, CO$_2$ to CO conversion, plasma-assisted combustion and plasma medicine.

Time-resolved electrical and optical measurements have been conducted to characterize the main features of the plasma discharge in the DBD microplasma reactor. Pulse energies per channel of about 1.9 $\mu$J at atmospheric pressure and 2.7 $\mu$J at 50 mbar have been calculated from the current and voltage time evolution. The discharge at low pressure is characterized by high vibrational temperatures (3980 K) and high reduced electric field strengths (2100 Td) compared to atmospheric pressure (3460 K and 500 Td), which indicates a higher electron energy at lower pressure. Based on natural luminosity images, all gas flowing through the reactor can be assumed to interact with the plasma at 50 mbar. At higher pressures, the emission concentrates at the channel walls and not all gas is in direct contact with the plasma. The distribution of the plasma-generated radicals, however, is unknown. The luminous part of the plasma is strongly non-thermal, with vibrational temperatures of about 4000 K, but the gas temperature remains only a few degrees above ambient.

In addition, we have noticed that the discharge emission intensity slightly decreases by increasing the air flow velocity through the channels at atmospheric pressure due to the lower pre-ionization level at higher velocities. Finally, a reactor lifetime study showed around 100 minutes lifetime at low pressure. Nevertheless, more effort is needed to increase the lifetime of the reactor in order to transfer it to real applications.


\section*{Acknowledgment}
A.Elkholy is supported by a Ph.D. grant from ministry of higher education, government of Egypt.

\section*{References }
\bibliographystyle{ieeetr}
\bibliography{ref}

\end{document}